\documentclass[superscriptaddress,nofootinbib,aps,twocolumn,prx]{revtex4-2}
\usepackage{typearea,physics,amsthm,bm,amsmath,amsfonts,natbib,qcircuit}
\usepackage[dvipdfmx]{graphicx}
\usepackage[colorlinks=true,
allcolors=blue,linktocpage=true]{hyperref}
\usepackage{algorithm,algcompatible,amsmath,mathtools}
\usepackage[caption=false]{subfig}
\algnewcommand\INPUT{\item[\textbf{Input:}]}%
\algnewcommand\OUTPUT{\item[\textbf{Output:}]}%
\algnewcommand\RETURN{\item[\textbf{return}]}%

\typearea{18}

\newcommand{\diamondnorm}[1]{|| #1 ||_{\diamond}}
\newcommand{\qswift}[1]{\mathcal{E}^{(#1)}}
\newcommand{\diamondDistance}[2]{
d_\diamond\left(#1, #2\right)
}

\newcommand{\sorting}[1]{f_{#1, k, N-k}}
\newcommand{\sortingDefault}[0]{\sorting{\sigma}}
\newcommand{\psiTwoJ}[0]{\mathcal{K}(r, \vec{\ell})}
\newcommand{\psiTwoJTilde}[0]{\tilde{\mathcal{K}}(r, \vec{\ell})}
\newcommand{\mixture}[0]{{M}}

\newcommand{\fig}[4]{
\begin{figure*}[ht]
\centering{
	\includegraphics[width=#1]{#2}
 }
	\caption{#3}
	\label{#4}
\end{figure*}
}
\newcommand{\figthree}[6]{
\begin{figure*}[ht]
\centering
	\subfloat[]{\includegraphics[width=#4]{#1}}
	\subfloat[]{\includegraphics[width=#4]{#2}}
	\subfloat[]{\includegraphics[width=#4]{#3}}
	\caption{#5}
	\label{#6}
\end{figure*}
}
\newcommand{\upi}{\text{i}} 
\newcommand{\upe}{\text{e}} 
\usepackage{xcolor}

\newcommand{\en}[0]{\mathcal{E}_N}

\newcommand{\swiftZero}[0]{\tilde{\mathcal{S}}_\ell^{(0)}}
\newcommand{\swiftOne}[0]{\tilde{\mathcal{S}}_\ell^{(1)}}

\newcommand{\chemUofT}{\affiliation{%
    Chemical Physics Theory Group, Department of Chemistry, University of Toronto, Toronto, Ontario, Canada}
    }
\newcommand{\csUofT}{\affiliation{%
    Department of Computer Science, University of Toronto, Toronto, Ontario, Canada}}

\newcommand{\vecInst}{\affiliation{%
    Vector Institute for Artificial Intelligence, Toronto, Ontario, Canada}}
    
\newcommand{\chemEngUofT}{\affiliation{%
    Department of Chemical Engineering \& Applied Chemistry, University of Toronto, Toronto, Ontario, Canada}}
\newcommand{\matSciUofT}{\affiliation{%
    Department of Materials Science \& Engineering, University of Toronto, Toronto, Ontario, Canada}}
\newcommand{\CIFAR}{\affiliation{%
    Lebovic Fellow, Canadian Institute for Advanced Research, Toronto, Ontario, Canada}}
\newcommand{\AIST}{\affiliation{%
    Research Center for Emerging Computing Technologies, National Institute of Advanced Industrial Science and Technology (AIST), 1-1-1 Umezono, Tsukuba, Ibaraki 305-8568, Japan}}
\newcommand{\Keio}{\affiliation{%
    Quantum Computing Center, Keio University, 3-14-1 Hiyoshi, Kohoku-ku, Yokohama, Kanagawa, 223-8522, Japan}}   

\newcommand{\knadd}[1]{
#1
}

\theoremstyle{plain}
\newtheorem{theorem}{Theorem}[section]
\newtheorem{lemma}[theorem]{Lemma}

\theoremstyle{definition}
\newtheorem{definition}[theorem]{Definition}

\usepackage{upgreek}

\begin{document}
\title{qSWIFT: High-order randomized compiler for Hamiltonian simulation
}
\author{Kouhei Nakaji}
\email{kohei.nakaji@utoronto.ca}
\chemUofT\AIST\Keio
\author{Mohsen Bagherimehrab}
\email{mohsen.bagherimehrab@utoronto.ca}
\chemUofT
\csUofT
\author{Al\'an Aspuru-Guzik}
\chemUofT\csUofT\vecInst\chemEngUofT\matSciUofT\CIFAR
\begin{abstract}    
Hamiltonian simulation is known to be one of the fundamental building blocks of a variety of quantum algorithms such as its most immediate application, that of simulating many-body systems to extract their physical properties. In this work, we present qSWIFT, a high-order randomized algorithm for Hamiltonian simulation. In qSWIFT, the required number of gates for a given precision is independent of the number of terms in Hamiltonian, while the systematic error is exponentially reduced with regards to the order parameter. In this respect, our qSWIFT is a higher-order counterpart of the previously proposed quantum stochastic drift protocol (qDRIFT), in which the number of gates scales linearly with the inverse of the precision required. We construct the qSWIFT channel and establish a rigorous bound for the systematic error quantified by the diamond norm. qSWIFT provides an algorithm to estimate given physical quantities using a system with one ancilla qubit, which is as simple as other product-formula-based approaches such as regular Trotter-Suzuki decompositions and qDRIFT. 
Our numerical experiment reveals that the required number of gates in qSWIFT is significantly reduced compared to qDRIFT. Particularly, the advantage is significant for problems where high precision is required; for example, to achieve a systematic relative propagation error of $10^{-6}$, the required number of gates in third-order qSWIFT is 1000 times smaller than that of qDRIFT. 
\end{abstract}
\maketitle
\tableofcontents

\section{Introduction}
\label{section:introduction}
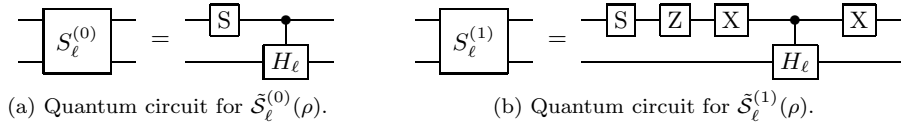
\begin{figure*}[ht]
\subfloat[Quantum circuit for $\swiftZero (\rho)$.]{
 ~\Qcircuit @C=1em @R=0.2em {
&\multigate{2}{S_\ell^{(0)}} &\qw &\lstick{} & \lstick{} & \gate{{\rm S}} & \ctrl{2} & \qw \\ 
&  && = & & &\\
&\ghost{S_\ell^{(0)}} &\qw &\lstick{}& \lstick{} & \qw & \gate{H_\ell} & \qw 
}
}~~~~~~~~
\subfloat[Quantum circuit for $\swiftOne (\rho)$.]{
\Qcircuit @C=1em @R=.2em {
&\multigate{2}{S_\ell^{(1)}} &\qw &\lstick{} & \lstick{}&\gate{{\rm S}}&\gate{{\rm Z}}&\gate{{\rm X}} &\ctrl{2} & \gate{{\rm X}}&\qw \\
&  && = & & &&&\\
&\ghost{S_\ell^{(1)}} &\qw &\lstick{}&\lstick{}&\qw & \qw & \qw &
\gate{H_\ell}&\qw & \qw
}
}
\caption{Quantum circuits for implementing the swift operators $\tilde{\mathcal{S}}_\ell^{(b)}  := S_\ell^{(b)} \rho S_\ell^{(b) \dagger}$ $(b \in \{0, 1\})$. 
The top line corresponds to an ancilla qubit, and the bottom line corresponds to the qubits in the system. \knadd{The ${\rm S}$ gate is defined by the operator $e^{\upi \pi/2} e^{\upi \sigma_z}$ with $\sigma_z$ is the Pauli-Z operator.}}
\label{fig:swiftChannel}
\end{figure*}

\begin{figure*}[ht]
\subfloat[A quantum circuit for qDRIFT with $N$ segments.]{
~\Qcircuit @C=.7em @R=1em {
&\qw & \gate{e^{\upi H_{\ell_1} \tau}} & \qw &
\gate{e^{\upi H_{\ell_2} \tau}} & \qw &\rstick{\cdots} & \lstick{} & \lstick{} & \lstick{} & \gate{e^{\upi H_{\ell_N} \tau}} & \qw
}
}{\label{fig:circuitQDrift}} \\
\subfloat[A swift circuit with $N$ segments, where two swift operators are performed in the $(r+1)$-th segment.]{
 ~\Qcircuit @C=.3em @R=.7em {
&\qw & \gate{{\rm H}} &\qw & \qw & \qw & \qw & \qw & \qw & \qw & \qw & \qw
& \qw & \multigate{1}{S_{\ell_{r+1}}^{(b_1)}} 
&\qw & \multigate{1}{S_{\ell_{r+1}^{\prime}}^{(b_2)}} & \qw & \qw & \qw & \qw & \qw
& \qw & \qw & \qw & \qw & \qw& \qw & \qw
 \\
&\qw & \gate{e^{\upi H_{\ell_1} \tau}} & \qw &
\rstick{\cdots} & \lstick{} & \lstick{} & \lstick{} & \lstick{} & \lstick{} & \lstick{} & \gate{e^{\upi H_{\ell_{r}} \tau}} & \qw 
&
\ghost{S_{\ell_{r+1}}^{(b_1)}}
&
\qw
&
\ghost{S_{\ell_{r+1}^{\prime}}^{(b_2)}}
&
\qw
&
\gate{e^{\upi H_{\ell_{r+2}} \tau}} & \qw &\rstick{\cdots} & \lstick{} & \lstick{} & \lstick{} &  
\lstick{} & \lstick{} & \lstick{} &\gate{e^{\upi H_{\ell_{N}} \tau}} & \qw
}
}{\label{fig:circuitQSWIFT}}
\caption{Examples of quantum circuits used for qDRIFT and qSWIFT.}
\label{fig:circuitQSWIFTALL}
\end{figure*}
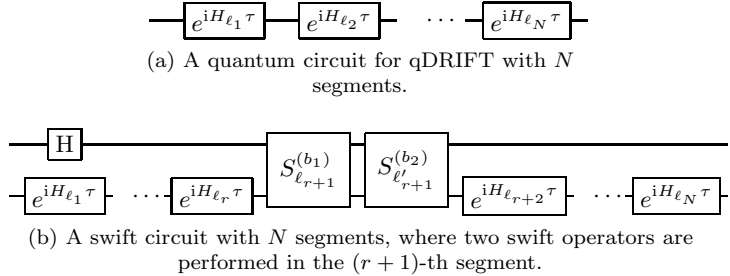

Hamiltonian simulation is a key subroutine of quantum algorithms for simulating quantum systems.
Given a Hamiltonian~$H=\sum_{\ell=1}^L h_\ell H_\ell$, where $h_\ell\geq 0$ and $L$ is the number of terms, the task in Hamiltonian simulation is to construct a quantum circuit that approximately emulate time evolution~$U(t):=\exp(-\text{i}Ht)$ of the system for time~$t$.
Several approaches have been established for this task.
The conventional approach uses the Trotter-Suzuki decompositions that provide a deterministic way for Hamiltonian simulation \cite{suzuki1990fractal,BAC+07,childs2021theory}.
The gate count of this approach scales at least linearly with the number of terms~$L$ in~$H$ \cite{childs2021theory}; we note that the gate count in \cite{BAC+07} scales at least quadratically with~$L$.
Although this scaling is formally efficient but is impractical for many applications of interest, particularly for the electronic-structure problem in quantum chemistry, where the number of terms in a Hamiltonian is prohibitively large.
An alternative approach is randomly permuting the order of terms in the Trotter-Suzuki decompositions~\cite{COS19}.
This randomized compilation provides a slightly better scaling for gate count over Ref.~\cite{BAC+07}, but the gate count still depends on the number of Hamiltonian terms quadratically.

The quantum stochastic drift protocol (qDRIFT) \cite{campbell2019random} is another randomized Hamiltonian simulation approach but is independent of the number of terms. 
In qDRIFT,  gates of the form~$\exp(-\text{i}H_\ell \tau)$ with small interval $\tau$ are applied randomly with a probability proportional to the strength~$h_\ell$ of the corresponding term in the Hamiltonian.
qDRIFT improves upon the Trotter-Suzuki approach in that its gate count is independent of~$L$ and~$\Lambda:=\max_\ell h_\ell$ (magnitude of the strongest term in the Hamiltonian), and instead depends on~$\lambda := \sum_{\ell=1}^L h_{\ell}$. 
However, qDRIFT has poor scaling with respect to the precision~$\varepsilon$ in contrast to that in the Trotter-Suzuki approach.
We note that there are other approaches to Hamiltonian simulation with asymptotically better performance as a function of various parameters~\cite{BC12,BCC+15,BCK15,LC17,LC19,GSL+19}.
Still, the approaches based on product formulae, e.g., Trotter-Suzuki decompositions and qDRIFT, are preferred for their superior performance in practice~\cite{CMN+18} and predominant usage in experimental implementations~\cite{BCC06,LHN+11,BLK+19} due to their simplicity and the fact that they do not require any ancilla qubits. From this perspective, we focus on an approach based on product formulae while having better gate scaling than the previous methods.

In this paper, we propose the {\it quantum swift protocol (\it qSWIFT)}\footnote{The code for the qSWIFT algorithm is available at \url{https://github.com/konakaji/qswift}.}, a high-order randomized algorithm having (i) better scaling with respect to the precision~$\varepsilon$ and (ii) the same scaling with respect to~$\lambda$ compared to qDRIFT. 
Specifically, the gate count of qSWIFT scales as~$\mathcal{O}((\lambda t)^2/\varepsilon^{\frac{1}{K}})$ with $K$ as the order-parameter while 
that of qDRIFT scales as~$\mathcal{O}((\lambda t)^2/\varepsilon)$. For example, with respect to the precision~$\varepsilon$, the gate count of qSWIFT scales as $\mathcal{O}(1/\sqrt{\varepsilon})$ for the second-order and as $\mathcal{O}(1/\varepsilon^{\frac{1}{3}})$ for the third-order.
Our qSWIFT algorithm shares its simplicity with the other approaches based on product formulae. It works in the system with one ancilla qubit (we refer to the qubits other than the ancilla qubit simply as the system qubits). We can construct all gate operations with $\exp(\upi H_{\ell} \tau)$ and the {\it swift operators} $\tilde{\mathcal{S}}_\ell^{b} := S_\ell^{(b)} \rho S_\ell^{(b) \dagger}$ where $S_\ell^{(b)}$ is a unitary transformation; the swift operators can be constructed if we can efficiently implement controlled-$H_\ell$ gates, as shown in Fig.~\ref{fig:swiftChannel}. 
In the case of qDRIFT, the entire time evolution is divided into segments, and a sampled time evolution $\exp(\upi H_\ell \tau)$ is performed in each segment (see Fig.~\ref{fig:circuitQDrift}). In qSWIFT, we utilize the {\it swift circuit} in addition to the circuit for qDRIFT. The swift circuit also has segments; in most segments, a sampled time evolution $\exp(\upi H_\ell \tau)$ is performed to the system qubits, but in the other segments, a sequence of the swift operators is performed (see Fig.~\ref{fig:circuitQSWIFT}). 
The number of swift operators is upper bounded by about twice the order parameter. 
Therefore, the qSWIFT algorithm can be performed with almost no additional resources compared to the qDRIFT.

We will now describe in more detail how qSWIFT is carried out. First, we build the qSWIFT channel that simulates the ideal time evolution. We then establish a bound for the distance between the qSWIFT channel and the ideal channel, quantified with the diamond norm, which exponentially decreases by increasing the order parameter. The established bound yields the desired scaling for the gate count of qSWIFT. It should be noted that the qSWIFT channel itself is not physical in the sense that it is not a completely-positive and trace-preserving~(CPTP) map. Nevertheless, we can employ the qSWIFT channel to develop a procedure for measuring a physical quantity of interest, i.e., computing the expectation value of some given observable that is exponentially more precise than the original qDRIFT with respect to the order parameter.

Our numerical analysis also reveals the advantage of our qSWIFT algorithm. 
We show the asymptotic behavior of qSWIFT by using electronic molecular Hamiltonians with the number of qubits $\sim 50$ and compare the performance with the other approaches based on the product formulae. Specifically, we compute the required number of gates to approximate the time evolution with the molecule Hamiltonians with a given systematic error $\varepsilon$. 
We show that the number of gates in the third-order version of qSWIFT is 10 times smaller than that of qDRIFT when $\varepsilon = 0.001$ for every time region. 
A significant reduction of the number of gates is observed when $\varepsilon = 10^{-6}$; the required number of gates in third-order (sixth-order) qSWIFT is $1,000~(10,000)$ times smaller than that of qDRIFT. We also simulate our qSWIFT algorithm and the other product formulae-based algorithms by using a quantum circuit simulator with the small-size (eight-qubits) molecular Hamiltonian. Its result is consistent with the result of the asymptotic behavior analysis. 

The rest of the paper is organized as follows. 
In Section~\ref{section:background}, we briefly review approaches for the Hamiltonian simulation based on the product formula. 
Section~\ref{section:secondOrder} and Section~\ref{section:higherOrder} are dedicated to proposing and analyzing our qSWIFT algorithm. In Section~\ref{section:secondOrder}, we introduce the way of constructing the second-order qSWIFT algorithm. Then we generalize the algorithm to the higher-order in Section~\ref{section:higherOrder}. 
In Section~\ref{section:numerics}, we validate our algorithm by numerical experiments. 
Finally, in Section~\ref{section:conclusion}, we conclude with some discussions.

\section{Background}
\label{section:background}
This section covers the key background pertinent to the following sections. 
We begin with a brief description of Hamiltonian simulation and Trotter-Suzuki formulae in Section~\ref{section:hs}.
Then we review the qDRIFT algorithm for Hamiltonian simulation in Section~\ref{section:qdrift}. 

\subsection{Hamiltonian simulation by Trotter-Suzuki formulae}
\label{section:hs}
We begin with a brief description of the Hamiltonian simulation.
For a given time-independent Hamiltonian of the form~$H=\sum_{\ell=1}^L h_\ell H_\ell$, where $h_\ell> 0$ and $H_\ell$ are Hermitian operators with $\norm{H_\ell}= 1$,  the task in Hamiltonian simulation is
to find a good approximation of the transformation 
\begin{equation}	
\label{eq:idealU}
\mathcal{U}(t): \rho \rightarrow U(t) \rho U^{\dagger}(t),
\end{equation}
with $U(t):=\upe^{\upi Ht}$ and $t$ as a real parameter. We assume we can efficiently implement each $\upe^{\upi H_\ell t^{\prime}}$ by quantum gates with $t^{\prime}$ as a real number. For example, if we decompose $H$ to the sum of the tensor products of the Pauli operators, we can efficiently implement each $\upe^{\upi H_\ell t^{\prime}}$.

The conventional approach for the Hamiltonian simulation is the Trotter-Suzuki decomposition \cite{BAC+07,suzuki1990fractal}.
In the first-order Trotter-Suzuki decomposition for Hamiltonian simulation, the entire simulation for time~$t$ is divided into $r$ segments of simulations for time $t/r$ as $U(t)=(U(t/r))^r$ and~$U(t/r)$ is approximated as $U(t/r)\approx U_\textsc{ts}^{(1)}(t/r)$~with
\begin{equation}
    U_\textsc{ts}^{(1)}(t):=\prod_{\ell=1}^L \upe^{\upi h_\ell H_\ell t},
\end{equation}
which yields the approximation
$U(t)\approx(U_\textsc{ts}^{(1)}(t/r))^r$
for the entire simulation.
The second-order Trotter-Suzuki decomposition is given by
$U(t)\approx(U_\textsc{ts}^{(2)}(t/r))^r$ with
\begin{equation}
    U_\textsc{ts}^{(2)}(t):=
    \prod_{\ell'=L}^1
    \upe^{\upi h_{\ell'} H_{\ell'} t/2}
    \prod_{\ell=1}^L
    \upe^{\upi h_\ell H_\ell t/2},
\end{equation}
which serves as the base case for the recursive formula
\begin{equation}
\begin{split}	
    &U^{(2k)}_\textsc{ts}(t) \\
    &:= \left[U^{(2k-2)}_\textsc{ts}(p_k t)\right]^2
    U^{(2k-2)}_\textsc{ts}((1-4p_{k})t) 
    \left[U^{(2k-2)}_\textsc{ts}(p_k t)\right]^2
\end{split}
\end{equation}
for the $2k^\text{th}$-order decomposition, where $p_k:=1/(4-4^{1/(2k-1)})$.

Let us discuss the Trotter-Suzuki decomposition in the channel representation. The $2k^\text{th}$-order Trotter-Suzuki channel
$\mathcal{U}^{(2k)}_\textsc{ts}(t):\rho \rightarrow U^{(2k)}_\textsc{ts}(t)\rho U^{(2k)}_\textsc{ts}(t)$
is used to approximate the channel $\mathcal{U}(\rho) $ as $\mathcal{U}(\rho)\approx (\mathcal{U}^{(2k)}_\textsc{ts}(t/r))^r$. For a given channel $\mathcal{C}$, we denote by $\mathcal{C}^{r^{\prime}}$ the $r^{\prime}$-repetition of $\mathcal{C}$.
Previous analytic work 
\knadd{\cite{BAC+07,childs2019faster,childs2018toward}} shows that 
\begin{equation}
\diamondnorm{\mathcal{U}(t) - (\mathcal{U}^{(2k)}_\textsc{ts}(t/r))^r}\leq \varepsilon    
\end{equation}
for
$r \in \mathcal{O}(\alpha L \Lambda t ( \alpha L \Lambda t/\varepsilon)^{1/2k})$ with $\alpha:=2\cdot5^{k-1}$, where $\diamondnorm{\cdot}$ is the diamond norm.
We note that $\alpha$ here is defined so that $r\alpha L$ is the number of gates used in the $2k^\text{th}$-order decomposition. 
Hence the gate count for the $2k^\text{th}$-order Trotter-Suzuki decomposition, denoted by $G_\textsc{ts}$, is
\begin{equation}
\label{eq:gatecount}
    G_\textsc{ts} = r \alpha L\in  O\left(\frac{\alpha^2 L^2 \Lambda t(\alpha L\Lambda t)^{\frac{1}{2k}}}{\varepsilon^{\frac{1}{2k}}}\right).
\end{equation}
Notice that the gate count approaches to $\mathcal{O}(L^2\Lambda t)$ by increasing the order parameter $2k$, but the prefactor scales exponentially with $2k$.
Because of this rapidly growing prefactor, Trotter-Suzuki decompositions of finite orders, typically second $(k=1)$ or fourth order $(k=2)$, are used in practice~\cite{campbell2019random}.

\knadd{
We note that \cite{childs2021theory} demonstrates that the gate-count scaling can be more rigorously bounded using the commutator bounds, where the upper bound is represented as the commutation relation.}
However, we use the gate-count scaling in Eq.~\eqref{eq:gatecount} to compare qSWIFT against qDRIFT \cite{campbell2019random}, particularly for comparing the numerical experiments in \cite{campbell2019random}.

\subsection{Hamiltonian simulation by qDRIFT}
\label{section:qdrift}
Developed by Campbell~\cite{campbell2019random},
qDRIFT is an algorithm for Hamiltonians simulation using a randomized procedure.
While the procedure is randomized, with many repetitions the evolution stochastically drifts towards the target unitary.
Specifically, the exact time evolution is approximated by 
$N$ repetitions of the qDRIFT channel $\mathcal{E}_N$ as $\mathcal{U}\approx \mathcal{E}_N^N$, with the qDRIFT channel defined as
\begin{equation}
\label{eq:en}
    \en (\rho) := \sum_{\ell=1}^L p_\ell \mathcal{T}_\ell(\rho), 
\end{equation}
where
\begin{equation}		
\mathcal{T}_\ell(\rho) = \text{e}^{\text{i} H_\ell \tau} \rho \text{e}^{-\text{i} H_\ell \tau},
\end{equation} 
is the unitary channel that we call the {\it time operator}, and
\begin{equation}
\label{eq:variables}
    p_\ell := h_\ell/\lambda,\quad 
    \lambda = \sum_\ell h_\ell, \quad
    \tau := \lambda t/N,
\end{equation}
are three variables used through the paper.
To realize the qDRIFT channel $\en$, 
the index $\ell$ is sampled according to the probability $p_\ell$ and the the quantum state $\rho$ is evolved through the channel associated with the operator $\text{e}^{\text{i}H_\ell \tau}$.
	
For evaluating the systematic error of the approximation, they define the exact short time evolution $\mathcal{U}_N$ as 
\begin{equation}
\label{eq:un}
	\mathcal{U}_N(\rho) := e^{\upi Ht/N} \rho e^{-\upi Ht/N}.  
\end{equation}
Then, they show 
\begin{equation}
	\diamondDistance{\mathcal{U}_N}{\mathcal{E}_N} \leq \frac{ 2(\lambda t)^2}{N^2} e^{2\lambda t/ N},  
\end{equation}
where the diamond distance is defined as
\begin{equation}
	\diamondDistance{\mathcal{U}^{\prime}}{\mathcal{E}^{\prime}} := \frac{1}{2}\diamondnorm{\mathcal{U}^{\prime} - \mathcal{E}^{\prime}}.
\end{equation}
They utilize the diamond distance $\diamondDistance{\mathcal{U}}{\mathcal{E}_N^N}$ as the measure of the systematic error. By using the subadditive feature of the diamond distance, they obtain the bound for the diamond distance as:
\begin{equation}
\begin{split}	
\diamondDistance{\mathcal{U}}{\mathcal{E}_N^N} 
	&\leq N \diamondDistance{\mathcal{U}_N}{\mathcal{E}_N} \\
	&\leq \frac{2(\lambda t)^2}{N} e^{2\lambda t/ N} \\
	&\in \mathcal{O}\left( \frac{(\lambda t)^2}{N} \right).
\end{split}
\end{equation} 
In other words, to reduce the systematic error within $\varepsilon$, we need to set $N \in \mathcal{O}((\lambda t)^2/\varepsilon)$.

\knadd{}

In most of the applications of the Hamiltonian simulation, what we have interests is computing the expectation value of an observable after applying the time evolution operator $\mathcal{U}$. Let us write the expectation value as 
\begin{equation}
\label{eq:def-q}
q := {\rm Tr}(Q \mathcal{U}(\rho_{\rm init})),	
\end{equation}
where $Q$ is an observable and $\rho_{\rm init}$ is an input quantum state. By using the qDRIFT algorithm, we can approximately compute the value of $Q$ as 
\begin{equation}	
\label{eq:def-q-1}
q^{(1)} := {\rm Tr}\left(Q \mathcal{E}_N^N(\rho_{\rm init}) \right),
\end{equation}
where the systematic error is bounded as
\begin{equation}
\label{eq:q-1-error}
	|q - q^{(1)}| \leq 2||Q||_{\infty} \diamondDistance{\mathcal{U}}{\mathcal{E}_N^N} 
 \in \mathcal{O}\left( ||Q||_{\infty}\left(\frac{(\lambda t)^2}{N}\right)\right). 
\end{equation}

\section{Second order qSWIFT}
\label{section:secondOrder}
In this section, we describe our second-order qSWIFT as a preparation for introducing the general high-order qSWIFT in Section~\ref{section:higherOrder}.
To elucidate our algorithm, we use a ``mixture function'' in our second and higher-order qSWIFT.
We begin by describing this function in Section~\ref{section:mixture}. 
Next, we construct the second-order qSWIFT channel and discuss its error bound in Section~\ref{sec:2ndqSWIFT}. Finally, in Section~\ref{section:secondOrderImpl}, we explain how to apply the constructed qSWIFT channel for computing physical quantities. 

\subsection{Mixture function}
\label{section:mixture}
In constructing our qSWIFT channels, we make use of a mixture function.
As a preparation, let us first define the following {\it sorting function}. 

\begin{definition}[Sorting function]
Let $S_N$ be the permutation group. 
For positive integers $k, N$ with $k < N$, let 
$\vec{\mathcal{A}}:=(\mathcal{A}_1,\ldots,\mathcal{A}_k)$ and
$\vec{\mathcal{B}}:=(\mathcal{B}_1,\ldots,\mathcal{B}_{N-k})$. 
We define the sorting function as 
	\begin{equation}
	\label{eq:sorting-function}
		\sortingDefault(\vec{\mathcal{A}}, \vec{\mathcal{B}}) := 
		\mathcal{X}_{\sigma(1)} \mathcal{X}_{\sigma(2)} \cdots \mathcal{X}_{\sigma(N)}, 
	\end{equation}
where $\sigma \in S_N$ and
\begin{equation}
\mathcal{X}_j = 
\begin{cases}
    \mathcal{A}_j & j\leq k,\\
    \mathcal{B}_{j-k} & j\geq k+1.
\end{cases}
\end{equation}
\end{definition}
\noindent The mixture function is defined by using the sorting function as follows.
\begin{definition}[Mixture function]
    For positive integers $k, N$ with $k < N$,
    let
$\vec{\mathcal{A}}:=(\mathcal{A}_1,\ldots,\mathcal{A}_k)$ and
$\vec{\mathcal{B}}:=(\mathcal{B}_1,\ldots,\mathcal{B}_{N-k})$.
Then we define the mixture function as
\begin{equation}
\label{eq:mixture}
    \mixture_{k,N-k}(\vec{\mathcal{A}},\vec{\mathcal{B}}) =
	\sum_{\sigma \in S^\text{sub}_{N,k}} 
       \sortingDefault(\vec{\mathcal{A}}, \vec{\mathcal{B}}),
\end{equation}
where the set~$S^\text{sub}_{N,k}$ is the subgroup of the permutation group~$S_N$ comprised of all elements~$\sigma\in S_N$ that satisfies the following condition:
if both $\mathcal{X}_{i}, \mathcal{X}_{j} \in \vec{\mathcal{A}}$ or $\in \vec{\mathcal{B}}$ then $\sigma(i) < \sigma(j)$ for any $i < j$. 
We remark that the number of elements in $S^\text{sub}_{N,k}$ is $\binom{N}{k}$.
\end{definition}

For simplicity, if elements of $\vec{\mathcal{B}}$ are identical, we denote the sorting and mixture functions as 
$\sortingDefault(\vec{\mathcal{A}},\mathcal{B})$ and  
$\mixture_{k,N-k}(\vec{\mathcal{A}},\mathcal{B})$, respectively.
In this case $\mathcal{X}_j=\mathcal{B}$ for $j\geq k + 1$.
Similarly, we use the notation $\sortingDefault(\mathcal{A},\mathcal{B})$ and $\mixture_{k,N-k}(\mathcal{A},\mathcal{B})$ if elements of $\vec{\mathcal{A}}$, and also elements of~$\vec{\mathcal{B}}$, are identical.
In this case, $\mathcal{X}_j=\mathcal{A}$ for $j\leq k$ and $\mathcal{X}_j=\mathcal{B}$ for $j\geq k+1$.

Note that the sorting function in Eq.~\eqref{eq:sorting-function} and mixture function in Eq.~\eqref{eq:mixture} are bilinear functions.
For example, if the $\ell$th element of~$\vec{\mathcal{A}}$ is a linear combination of elements of another vector~$\vec{\mathcal{F}}$, i.e.,
if~$\mathcal{A}_\ell=\sum_n c_n \mathcal{F}_n$ for~$c_n\in\mathbb{C}$, then we have
\begin{equation}
\begin{split}	
    &\mixture_{k,N-k}\left((\mathcal{A}_1,\ldots,\mathcal{A}_{\ell-1},\sum_n c_n \mathcal{F}_n, \mathcal{A}_{\ell+1},\ldots,\mathcal{A}_k),\vec{\mathcal{B}}\right) \\
    &= \sum_n c_n
    \mixture_{k,N-k}\left((\mathcal{A}_1,\ldots,\mathcal{A}_{\ell-1},\mathcal{F}_n, \mathcal{A}_{\ell+1},\ldots,\mathcal{A}_k),\vec{\mathcal{B}}\right).
\end{split}
\end{equation}
In general, if~$\mathcal{A}_\ell=\sum_{n_\ell} c_{n_\ell} \mathcal{F}_{n_\ell}$ for any~$\ell$, then the identity
\begin{equation}
\begin{split}	
\label{eq:vecP}
      \mixture_{k,N-k}\left(\vec{\mathcal{A}},\vec{\mathcal{B}}\right)
    &= \sum_{n_1} c_{n_1}\sum_{n_2} c_{n_2}\cdots \sum_{n_k} c_{n_k} \\
    &~~~~~~\mixture_{k,N-k}\left((\mathcal{F}_{n_1},\ldots,\mathcal{F}_{n_k}),\vec{\mathcal{B}}\right)
\end{split}
\end{equation}
holds.

\subsection{Second-order qSWIFT channel}
\label{sec:2ndqSWIFT}
To construct the qSWIFT channel, let us define
\begin{align}
\label{eq:lj-def}
    &\mathcal{L}_\ell(\rho) := \upi [H_\ell, \rho], \\
    \label{eq:l-def}
    &\mathcal{L}(\rho):= \frac{\upi}{\lambda}[H, \rho] = \sum_\ell p_\ell \mathcal{L}_\ell(\rho),
\end{align}
where the variables $\lambda, p_\ell$ and $ \tau$ are defined in Eq.~\eqref{eq:variables}.
We then have
\begin{align}
    &\label{eq:unExpand}
    \mathcal{U}_N = \upe ^{\mathcal{L}\tau } = \mathbb{I} + \tau\mathcal{L} + \Delta^{(2)}\mathcal{U}_N,\\
    &\label{eq:enExpand}
    \mathcal{E}_N = \sum_\ell p_\ell \upe^{\mathcal{L}_\ell\tau} = \mathbb{I} + \tau\mathcal{L}+ \Delta^{(2)}\mathcal{E}_N,
\end{align}
for the ideal time-evolution channel in Eq.~\eqref{eq:un}
and the qDRIFT channel in Eq.~\eqref{eq:en},
where 
\begin{align}	
&\Delta^{(k)}\mathcal{U}_N = \sum_{n=k}^{\infty} \frac{\tau^{n}}{n!} \mathcal{L}^n, \\
&\Delta^{(k)}\mathcal{E}_N = \sum_{n=k}^{\infty} \frac{\tau^{n}}{n!} \sum_{\ell=1}^L p_\ell  \mathcal{L}_\ell^n.
\end{align}
Let $\Delta_k:=\Delta^{(k)}\mathcal{U}_N-\Delta^{(k)}\mathcal{E}_N$, then
\begin{equation}
\label{eq:Deltak}
    \Delta_k = \sum_{n=k}^\infty \frac{\tau^n}{n!}\mathcal{L}^{(n)},
    \quad
    \mathcal{L}^{(n)}:=\mathcal{L}^n-\sum_{\ell=1}^{L}p_\ell\mathcal{L}^n_\ell.
\end{equation}
Using the definition of $\Delta_k$ and Eqs.~\eqref{eq:unExpand} and~\eqref{eq:enExpand}, we have $\mathcal{U}_N = \mathcal{E}_N + \Delta_2$ which we use to expand 
$\mathcal{U}=\mathcal{U}_N^N$ as
\begin{equation}
\begin{split}
    \mathcal{U}
    &= (\mathcal{E}_N+\Delta_2)^N\\
    &= \mathcal{E}^N_N + \sum_{k=1}^{N} \mixture_{k,N-k}\left(\Delta_2, \mathcal{E}_N\right)\\
    &= \mathcal{E}^N_N
    + \frac{\tau^2}{2} \mixture_{1,N-1}\left(\mathcal{L}^{(2)}, \mathcal{E}_N\right)
    + \mixture_{1,N-1}\left(\Delta_3, \mathcal{E}_N\right) \\
    &~~~+\sum_{k=2}^{N} \mixture_{k,N-k}\left(\Delta_2, \mathcal{E}_N\right),
\end{split}	
\end{equation}
where we used~$\Delta_2=(\tau^2/2)\mathcal{L}^{(2)}+\Delta_3$ and linearity of~$\mixture_{1,N-1}$ to obtain the last equality.
Let us denote the first two terms as
\begin{equation}
\label{eq:2ndqSWIFT}
    \qswift{2}:= \mathcal{E}^N_N
    + \frac{\tau^2}{2} \mixture_{1,N-1}\left(\mathcal{L}^{(2)}, \mathcal{E}_N\right).
\end{equation}
We refer to~$\qswift{2}$ as the \textit{second-order qSWIFT} channel.
In the following lemma, we provide a bound for the error in approximating the ideal channel~$\mathcal{U}$ in Eq.~\eqref{eq:idealU} by the second-order qSWIFT channel~$\qswift{2}$,
where the error is quantified as the diamond norm of their difference.
\begin{lemma}
\label{lemma:2ndqSWIFTErr}
Let~$\mathcal{U}$ be the ideal channel in Eq.~\eqref{eq:idealU} and let~$\qswift{2}$ be the second-order qSWIFT channel in Eq.~\eqref{eq:2ndqSWIFT}.
Then, in the region $\lambda t \geq 1$,
\begin{equation}
\label{eq:2ndqSWIFTErr}
	\diamondDistance{\mathcal{U}}{\qswift{2}}
	\in \order{\left(\frac{(\lambda t)^2}{N}\right)^2},
\end{equation}
\knadd{provided} $N \leq 2\sqrt{2}e (\lambda t)^2$.
\end{lemma}
\noindent
We provide the proof in Appendix~\ref{section:secondErrorBound}.
Invoking this lemma for the reasonable parameter region $\lambda t \geq 1$, 
if $N\in \order{(\lambda t)^2/\sqrt{\varepsilon}}$ for $\varepsilon>0$, then~$\diamondDistance{\mathcal{U}}{\qswift{2}}\leq\varepsilon$.
This result provides a quadratic improvement over the original qDRIFT with respect to~$\varepsilon$.  

\subsection{Implementation of the second-order qSWIFT channel}
\label{section:secondOrderImpl}
The second-order qSWIFT channel we constructed is not a physical channel as it is not a CPTP map. Therefore, we cannot directly implement the qSWIFT channel itself. However, in most applications of the Hamiltonian simulation, our interest is in computing physical quantities, i.e., the expectation value of an observable after applying the time evolution operator as described in Section~\ref{section:qdrift}. Thus, in the following, we focus on how to compute $q^{(2)} := {\rm Tr}(Q \qswift{2}(\rho_{\rm init}))$, for a given observable $Q$ and the input $\rho_{\rm init}$. 
By using $q^{(2)}$, the systematic error is reduced to 
\begin{equation}
|q - q^{(2)}| \leq 2||Q||_{\infty} \diamondDistance{\mathcal{U}}{\qswift{2}}
 \in \mathcal{O}\left( ||Q||_{\infty}\left(\frac{(\lambda t)^2}{N}\right)^2\right),
\end{equation}
where we use \eqref{eq:2ndqSWIFTErr}, 
which is the quadratic improvement in terms of $(\lambda t)^2/N$ from the one in qDRIFT \eqref{eq:q-1-error}.

Here we provide a way of computing $q^{(2)}$ by quantum circuits. 
We can expand $q^{(2)}$ as 
\begin{equation}
\label{eq:q2}
	q^{(2)} = q^{(1)} + 
	 \delta q, 
\end{equation}
where
\begin{equation}
\label{eq:q2-expansion}
\delta q = 
\frac{\tau^2}{2} {\rm Tr}\left(Q\mixture_{1,N-1}\left(\mathcal{L}^{(2)}, \mathcal{E}_N\right) (\rho_{\rm init})\right). 
\end{equation}
For computing $q^{(1)}$, we just need to apply the original qDRIFT channel and compute the expectation value. Thus, we focus on how to compute $\delta q$ in the following. 
More specifically, we will show that $\delta q$ is computable by using the swift circuits, composed of the time evolution $\exp\left(\upi H_\ell \tau \right)$ and the swift operators shown in Fig.~\ref{fig:swiftChannel}. 

By using 
\begin{equation}
	\mixture_{1,N-1}\left(\mathcal{L}^{(2)}, \mathcal{E}_N\right) 
	= \sum_{r=0}^{N-1} \mathcal{E}_N^{N-1-r} \mathcal{L}^{(2)} \mathcal{E}_N^r, 
\end{equation}
which can be derived from the definition \eqref{eq:mixture}, we obtain
\begin{equation}
\label{eq:del-q-two}	
	\delta q = \frac{\tau^2}{2} \sum_{r=0}^{N-1} \delta q_r, 
\end{equation}
where $\delta q_r := {\rm Tr}\left(Q \mathcal{E}_N^{N-1-r} \mathcal{L}^{(2)} \mathcal{E}_N^r (\rho_{\rm init})\right)$.
Now let us move on to the evaluation of $\delta q_r$. 
We  transform $\delta q_r$ as 
\begin{equation}
\begin{split}
	\delta q_r &= {\rm Tr}\left(Q \mathcal{E}_N^{N-1-r} \mathcal{L}^{(2)} \mathcal{E}_N^r (\rho_{\rm init}) \right), \\
	&= 
	\sum_{\ell,k=1}^L p_\ell p_k
	{\rm Tr}\left(Q \mathcal{E}_N^{N-1-r} 
	\mathcal{L}_\ell \mathcal{L}_k \mathcal{E}_N^r (\rho_{\rm init}) \right)\\
	&~~~- \sum_{\ell=1}^L p_\ell {\rm Tr}\left(Q \mathcal{E}_N^{N-1-r} \mathcal{L}_\ell^2 \mathcal{E}_N^r (\rho_{\rm init}) \right),
\end{split}	
\end{equation}
where we use \eqref{eq:Deltak} in the second equality.
Let two probability distributions be
\begin{align}
 \label{eq:prob-expand-2}
	P_0^{(2)} (\vec{\ell}) = p_{\ell_2} p_{\ell_1}, 
	~P_1^{(2)} (\vec{\ell}) = 
	\begin{cases}
			p_\ell & \ell_1 = \ell_2 = \ell\\
			0    & \ell_1 \neq \ell_2 
	\end{cases}, 
\end{align}
where the input $\vec{\ell}$ is a vector with two elements $(\ell_1, \ell_2)$. 
Also, let 
\begin{equation}
\label{eq:twoL}
	\mathcal{L}_2(\vec{\ell}) := 
	\mathcal{L}_{\ell_2} \mathcal{L}_{\ell_1}.
\end{equation}
Then it holds 
\begin{equation}
\label{eq:delQLImproved}
	\delta q_r := \sum_{s=0}^{1}(-1)^s \sum_{\vec{\ell}}P_s^{(2)}(\vec{\ell})
	{\rm Tr}\left(Q
	\psiTwoJ (\rho_{\rm init})
	\right),
\end{equation}
where we define a channel $\psiTwoJ$ as
\begin{equation}
	\psiTwoJ := \mathcal{E}_N^{N-1-r} 
	\mathcal{L}_2(\vec{\ell})
	 \mathcal{E}_N^r.
\end{equation}
 
 To evaluate each term of \eqref{eq:delQLImproved}, we utilize a system with one ancilla qubit. 
Let us write the density matrix for a given system with one ancilla qubit as the matrix form:
\begin{equation}
\begin{split}	
	&\left(
	\begin{array}{cc}
		\rho_{00} & \rho_{01} \\
		\rho_{10} & \rho_{11}
	\end{array}
	\right) 
	 \\
	&:= |0\rangle \langle 0| \otimes \rho_{00} + |0\rangle \langle 1| \otimes \rho_{01}
	+ |1\rangle \langle 0| \otimes \rho_{10} + |1\rangle \langle 1| \otimes \rho_{11}.
\end{split}
\end{equation}
We define the operation of a quantum channel $\psiTwoJTilde$
that transforms the initial state 
\begin{equation}
    \tilde{\rho}_{\rm init} := |+\rangle\langle +| \otimes \rho_{\rm init} =
    \left(
	\begin{array}{cc}
		\rho_{\rm init}/2 & \rho_{\rm init}/2 \\
		\rho_{\rm init}/2 & \rho_{\rm init}/2
	\end{array}
	\right)
\end{equation}
into a final state as
\begin{equation}
\begin{split}	
\label{eq:jk-l}
\tilde{\rho}_{\rm init}
	\xmapsto{\psiTwoJTilde}
		\left(
	\begin{array}{cc}
		\cdot & \tfrac{1}{2} \psiTwoJ(\rho_{\rm init}) \\
		\tfrac{1}{2} \psiTwoJ (\rho_{\rm init}) & \cdot
	\end{array}
	\right),  
\end{split}
\end{equation}
where the dot [.] in the diagonal element denotes a matrix we do not have interest in now. 
Then it holds 
\begin{equation}
\label{eq:obs-equality}
{\rm Tr}(Q\psiTwoJ(\rho_{\rm init})) =  {\rm Tr}\left(\tilde{Q} \psiTwoJTilde(\tilde{\rho}_{\rm init}) \right),
\end{equation}
and consequently,
\begin{equation}
\label{eq:delQLAncilla}
		\delta q_r := \sum_{s=0}^{1}(-1)^s \sum_{\vec{\ell}}P_s^{(2)}(\vec{\ell})
	{\rm Tr}\left(\tilde{Q}
	\psiTwoJTilde (\tilde{\rho}_{\rm init})
	\right),
\end{equation}
where 
\begin{equation}
\tilde{Q} = X \otimes Q	,
\end{equation}
with $X$ as the Pauli-X observable for the ancilla qubit.
Therefore, we can evaluate $\delta q_r$ if we implement the channel $\psiTwoJTilde$ by using quantum circuits.

We can specify the channel $\psiTwoJTilde$ as follows:
\begin{equation}
\label{eq:tilde-psi}
\psiTwoJTilde := 
	\tilde{\mathcal{E}}_N^{N-1-r} \tilde{\mathcal{L}}_{\ell_2} \tilde{\mathcal{L}}_{\ell_1} \tilde{\mathcal{E}}_N^{r}, 
\end{equation}
where we define 
$\tilde{\mathcal{E}}_N := {\bf 1} \otimes \mathcal{E}_N$ and $\vec{\ell} = (\ell_1, \ell_2)$, and where  
the operation of the channel $\tilde{\mathcal{L}}_\ell$ is specified for the input having the identical non-diagonal elements as follows:
\begin{equation}
\label{eq:tilde-lj}
\left(
	\begin{array}{cc}
		\cdot & \rho \\
		\rho & \cdot
	\end{array}
\right) \xmapsto{\tilde{\mathcal{L}}_\ell}
\left(
	\begin{array}{cc}
		\cdot & \mathcal{L}_\ell(\rho) \\
		\mathcal{L}_\ell(\rho) & \cdot
	\end{array}
\right).  
\end{equation}
The channel $\tilde{\mathcal{L}}_\ell$ can be written as the sum of two swift operators 
$\swiftZero$ and $\swiftOne$ introduced in Fig.~\ref{fig:swiftChannel} as 
\begin{equation}
\label{eq:l-sum}
	\tilde{\mathcal{L}}_\ell = \swiftZero + \swiftOne, 
\end{equation}
which is easily checked by using
\begin{equation}
\begin{split}	
 \left(
\begin{array}{cc}
		\cdot & \rho \\
		\rho & \cdot
	\end{array}
\right)  \xmapsto{\swiftZero}
\left(
\begin{array}{cc}
		\cdot & -\upi \rho H_\ell \\
		\upi H_\ell \rho & \cdot
	\end{array}
\right), \\
 \left(
\begin{array}{cc}
		\cdot & \rho \\
		\rho & \cdot
	\end{array}
\right)  \xmapsto{\swiftOne}
\left(
\begin{array}{cc}
		\cdot & \upi H_\ell \rho  \\
		-\upi \rho H_\ell  & \cdot
	\end{array}
\right).
\end{split}
\end{equation}
It can be pedagogically shown that the channel \eqref{eq:tilde-psi} reproduces the transformation in \eqref{eq:jk-l}; with $\rho^{\prime}_{\rm init} = \rho_{\rm init}/2$, it holds 
\begin{equation}
\begin{split}	
	\left(
	\begin{array}{cc}
		\rho^{\prime}_{\rm init} & \rho^{\prime}_{\rm init} \\
		\rho^{\prime}_{\rm init} & \rho^{\prime}_{\rm init}
	\end{array}
	\right) 
	&\xmapsto{\tilde{\mathcal{E}}_N^{r}}
	\left(
	\begin{array}{cc}
		\mathcal{E}_N^{r}(\rho^{\prime}_{\rm init}) & \mathcal{E}_N^{r}(\rho^{\prime}_{\rm init}) \\
		\mathcal{E}_N^{r}(\rho^{\prime}_{\rm init}) & \mathcal{E}_N^{r}(\rho^{\prime}_{\rm init})
	\end{array}
	\right) \\ 
	&\xmapsto{\tilde{\mathcal{L}}_{\ell_1}} 
	\left(
	\begin{array}{cc}
		\cdot & \mathcal{L}_{\ell_1} \mathcal{E}_N^{r}(\rho^{\prime}_{\rm init}) \\
		\mathcal{L}_{\ell_1}  \mathcal{E}_N^{r}(\rho^{\prime}_{\rm init}) & \cdot
	\end{array}
	\right) \\
		&\xmapsto{\tilde{\mathcal{L}}_{\ell_2}} 
	\left(
	\begin{array}{cc}
		\cdot & \mathcal{L}_{\ell_2} 
		\mathcal{L}_{\ell_1} \mathcal{E}_N^{r}(\rho^{\prime}_{\rm init}) \\
		\mathcal{L}_{\ell_2}
		\mathcal{L}_{\ell_1}
		\mathcal{E}_N^{r}(\rho^{\prime}_{\rm init}) & \cdot
	\end{array}
	\right) \\
	&\xmapsto{\tilde{\mathcal{E}}_N^{N-1-r}}
		\left(
	\begin{array}{cc}
		\cdot & \psiTwoJTilde (\rho^{\prime}_{\rm init}) \\
		\psiTwoJTilde (\rho^{\prime}_{\rm init}) & \cdot
	\end{array}
	\right).  
\end{split}
\end{equation}

By substituting \eqref{eq:l-sum} to \eqref{eq:tilde-psi}, we obtain 
\begin{equation}
\label{eq:tilde_psi}
\psiTwoJTilde = \sum_{b_1, b_2 =0}^1  \tilde{\mathcal{E}}_N^{N-1-r} \tilde{\mathcal{S}}_{\ell_2}^{(b_2)}\tilde{\mathcal{S}}_{\ell_1}^{(b_1)} \tilde{\mathcal{E}}_N^{r}.
\end{equation}
Finally, from \eqref{eq:del-q-two}, \eqref{eq:delQLAncilla}, and \eqref{eq:tilde_psi}, 
\begin{equation}
\begin{split}
\label{eq:del-q-evaluatable}
	\delta q &= \frac{\tau^2}{2} 
    \sum_{s=0}^{1}(-1)^s 
	\sum_{b_1, b_2 =0}^1
	\sum_{r=0}^{N-1} \sum_{\vec{\ell}} \\
	&~~~~~~~~P_s^{(2)}(\vec{\ell})
	{\rm Tr}\left(\tilde{Q}
	\tilde{\mathcal{E}}_N^{N-1-r} \tilde{\mathcal{S}}_{\ell_2}^{(b_2)}\tilde{\mathcal{S}}_{\ell_1}^{(b_1)} \tilde{\mathcal{E}}_N^{r} (\rho_{\rm init})
	\right), \\
	&= \frac{(\lambda t)^2}{2N} \sum_{s=0}^1(-1)^s \sum_{b_1, b_2=0}^1 \delta q(s, b_1, b_2), 
\end{split}	
\end{equation}
where in the second equality, we define 
\begin{equation}
\begin{split}	
\label{eq:q2SB}
	 &\delta q(s, b_1, b_2) \\
	 &:= 
	 \frac{1}{N}
	 \sum_{r=0}^{N-1} 
	\sum_{\vec{\ell}}P_s^{(2)}(\vec{\ell})
	{\rm Tr}\left(\tilde{Q}
	\tilde{\mathcal{E}}_N^{N-1-r} \tilde{\mathcal{S}}_{\ell_2}^{(b_2)}\tilde{\mathcal{S}}_{\ell_1}^{(b_1)} \tilde{\mathcal{E}}_N^{r} (\tilde{\rho}_{\rm init})
	\right).
\end{split}
\end{equation}
We can evaluate \eqref{eq:q2SB} using Monte Carlo sampling. 
To this end, let
\begin{equation} P_\textsc{mdrift}(\vec{k}, n)
= p_{k_1} p_{k_2}\cdots p_{k_n}
\end{equation}
be the probability distribution for product of multiple qDRIFT probability distributions, where $n \in \mathbb{Z}^+$ specifies the number of qDRIFT distributions and $k_j$, for each $j$, goes from 1 to $L$. Then 
\begin{equation}
	\tilde{\mathcal{E}}_N^{n} 
	= \sum_{k_1, k_2, \cdots k_{n}}^L 
	P_\textsc{mdrift}(\vec{k}, n)
	\tilde{\mathcal{T}}_n({\vec{k}}),  
\end{equation}
where $\tilde{\mathcal{T}}_n({\vec{k}}) $ is the unitary channel defined as the sequence of the time operators:
\begin{equation}
	\tilde{\mathcal{T}}_n({\vec{k}}) :=  
	({\bf 1} \otimes \mathcal{T}_{k_{n}}
	\cdots \mathcal{T}_{k_2} \mathcal{T}_{k_{1}}
	).
\end{equation}
We obtain an unbiased estimator of  $\delta q(s, b_1, b_2)$ as follows: 
\begin{enumerate}
	\item Sample $\ell$ uniformly from $\{0, 1, \cdots N-1\}$.
	\item With probability $P_s^{(2)}(\vec{\ell})$ sample $\vec{\ell} = (\ell_1, \ell_2)$. With probability $P_\textsc{mdrift}(\vec{k}, r)$ and $P_\textsc{mdrift}(\vec{k}^{\prime}, N-1-r)$, sample $\vec{k}$ and $\vec{k}^{\prime}$.
	\item Estimate 
	\begin{equation}
	\label{eq:secondCircuit}
	{\rm Tr}
	\left(\tilde{Q} 
	\tilde{\mathcal{T}}_{ N-1-r}(\vec{k}^{\prime})
	\tilde{\mathcal{S}}^{(b_2)}_{\ell_2} \tilde{\mathcal{S}}^{(b_1)}_{\ell_1} \tilde{\mathcal{T}_{r}}(\vec{k})(\tilde{\rho}_{\rm init}) \right) 
	\end{equation}
	with $N_{\rm shot}$ measurements and set the resulting value to $\delta \hat{q}(s, b_1, b_2)$, which is an unbiased estimator of $\delta q(s, b_1, b_2)$. The value \eqref{eq:secondCircuit} can be evaluated by applying a swift circuit composed of the time evolution and the swift operators and estimating the expectation value of $\tilde{Q}$ by measurements. 
\end{enumerate}
We repeat the above process $N_{\rm sample}$ times, 
and the estimate of $\delta q(s, b_1, b_2)$ is computed as the sample average. By substituting each estimate of $\delta q(s, b_1, b_2)$ to \eqref{eq:q2SB}, we obtain the estimate of $\delta q$ as $\delta \hat{q}$. It should be noted that the swift
circuit for evaluating each term of \eqref{eq:secondCircuit} has the structure that two swift operators are tucked in between $N-1$ time operators as in Fig.~\ref{fig:circuitQSWIFT}.
Therefore, the number of gates in the swift circuit is 
almost the same as the original qDRIFT that requires $N$ time operators.


\section{Higher-order qSWIFT}
\label{section:higherOrder}
In this section, we generalize the second-order qSWIFT channel introduced in Section~\ref{sec:2ndqSWIFT} to an arbitrary high-order channel.
First we construct the higher-order qSWIFT channel and discuss the error bound in Section~\ref{section:higherOrderChannel}. 
Then, in Section~\ref{section:higherOrderImpl}, we construct an algorithm to apply the qSWIFT channel for computing physical quantities. 
\subsection{Higher-order qSWIFT channel}
\label{section:higherOrderChannel}
To construct a high-order qSWIFT channel, we retain higher orders of~$\tau$ in the right-hand-side of
\begin{equation}
    \mathcal{U}=(\mathcal{E}_N+\Delta_2)^N
    = \mathcal{E}^N_N + \sum_{k=1}^{N} \mixture_{k,N-k} (\Delta_2,\mathcal{E}_N),
\end{equation}
where $\mixture_{k,N-k}$ is the mixture function defined in Eq~\eqref{eq:mixture}.
We note that~$\Delta_2$ is a linear combination of~$\mathcal{L}^{(n)}$ as per Eq.~\eqref{eq:Deltak}.
Thus by Eq.~\eqref{eq:vecP} we obtain
\begin{equation}
\begin{split}
    &\mixture_{k,N-k} (\Delta_2,\mathcal{E}_N) \\
    &= \sum_{n_1=2}^{\infty} \frac{\tau^{n_1}}{n_1!}
    \sum_{n_2=2}^{\infty} \frac{\tau^{n_2}}{n_2!}
    \cdots
    \sum_{n_k=2}^{\infty} \frac{\tau^{n_k}}{n_k!} \\
    &~~~~~~~~~~\mixture_{k,N-k}
    \left(
    \left(\mathcal{L}^{(n_1)},\ldots,\mathcal{L}^{(n_k)}\right)
    ,\mathcal{E}_N
    \right),\\
    &=\sum_{n_1,n_2,\ldots,n_k=2}^{\infty}
    \frac{\tau^{\sum_{j=1}^k n_j}}{n_1!n_2!\cdots n_k!}
    \sum_{\xi=2}^{\infty}\updelta\left[\xi, \sum_{\ell=1}^k n_{\ell}\right] \\
    &~~~~~~~~~~\mixture_{k,N-k}
    \left(
    \left(\mathcal{L}^{(n_1)},\ldots,\mathcal{L}^{(n_k)}\right)
    ,\mathcal{E}_N
    \right), \\
    &= \sum_{\xi=2}^{\infty}\; \tau^\xi \sum_{n_1,n_2,\ldots,n_k=2}^{\xi}
    \frac{1}{n_1!n_2!\cdots n_k!}
    \updelta\left[\xi, \sum_{j=1}^k n_j\right] \\
    &~~~~~~~~~~\mixture_{k,N-k}
    \left(
    \left(\mathcal{L}^{(n_1)},\ldots,\mathcal{L}^{(n_k)}\right)
    ,\mathcal{E}_N
    \right),
\end{split}	
\end{equation}
where $\updelta[i,j]$ here is the Kronecker~$\updelta$. To show the second equality, we use the identity 
\begin{equation}
   \sum_{\xi = 2}^{\infty} \updelta\left[\xi, \sum_{j=1}^k n_{j} \right] = 1,
\end{equation}
which holds for a fixed set of integers $\{n_j \}_{j=1}^k$ with $n_{j}\geq 2$.
In the last equality, we use the fact that the Kronecker~$\updelta$ is zero if any of $n_1, n_2\cdots n_k$ is larger than $\xi$. 
Truncating the upper limit of $\xi$ yields a high-order qSWIFT channel.
Specifically, we define the high-order qDRIFT as follows.

\begin{definition}[Higher-order qSWIFT]
    We define $K$-th order qSWIFT channel ($K\leq N$) as
    \begin{equation}
    \begin{split}    	
    \label{eq:qSWIFTChannel}
        &\qswift{K} \\
        &:= \mathcal{E}_N^N
        +  \sum_{\xi=2}^{2K-2} \tau^\xi \sum_{k=1}^N \sum_{n_1,\ldots,n_k=2}^\xi 
        \frac{1}{n_1!n_2!\cdots n_k!} 
         \updelta\left[\xi, \sum_{j=1}^k n_j\right] \\
        &~~~~~~~~~~~~~~~~~~~\mixture_{k,N-k}
        \left(
        \left(\mathcal{L}^{(n_1)},\ldots,\mathcal{L}^{(n_k)}\right)
        ,\mathcal{E}_N
        \right).
    \end{split}
    \end{equation}
\end{definition}
Here we note that the upper limit $N$ in the second summation can be replaced with $K$ for $K\leq N$ because the terms with $k>K$ become zero by the Kronecker $\updelta$.
We remark that setting $K=2$ yields the second-order qSWIFT channel in Eq.~\eqref{eq:2ndqSWIFT}. 
Also, by setting $K=1$, we reproduce the qDRIFT channel. 

We now provide a bound for the error in approximating the ideal channel~$\mathcal{U}$ in Eq.~\eqref{eq:idealU} by the high-order qSWIFT channel~$\qswift{K}$ in the following lemma.

\begin{lemma}
\label{lemma:SWIFTErr}
Let~$\mathcal{U}$ be the ideal channel in and let~$\qswift{K}$ be the $K$-th order qSWIFT channel.
Then, in the region $\lambda t \geq 1$,
\begin{equation}
\label{eq:kqSWIFTErr}
	\diamondDistance{\mathcal{U}}{\qswift{K}}
	\in \order{\left(\frac{(\lambda t)^2}{N}\right)^K}.
\end{equation}
as far as $N \leq 2\sqrt{2}e (\lambda t)^2$.
\end{lemma}
\noindent The proof is given in Appendix~\ref{section:higherErrorBound}. 
As in the case of the second order, 
for $\lambda t \geq 1$ and $\varepsilon>0$, 
if $N\in \order{(\lambda t)^2/\sqrt{\varepsilon}}$, then~$\diamondDistance{\mathcal{U}}{\qswift{K}}\leq\varepsilon$.

\subsection{Implementation of the higher-order qSWIFT channel}
\label{section:higherOrderImpl}
As we discuss in Section~\ref{section:secondOrderImpl}, the qSWIFT channel is not a physical channel, but we can apply it to computing the physical quantities. Specifically, we discuss how to compute 
\begin{equation}
q^{(K)} := {\rm Tr}\left(Q\qswift{K}\right). 
\end{equation}
Then, the systematic error is bounded as 
\begin{equation}
\begin{split}	
|q - q^{(K)}| 
&\leq2 ||Q||_{\infty} \diamondDistance{\mathcal{U}}{\qswift{K} }\\
&\in \mathcal{O}\left( ||Q||_{\infty}\left(\frac{(\lambda t)^2}{N}\right)^{K}\right), 
\end{split}
\end{equation}  
which can be exponentially small with the order parameter. 

Here, we provide the way to compute $q^{(K)}$. 
We can expand $q^{(K)}$ as 
\begin{align}	
\label{eq:allOrder}
q^{(K)}	&= q^{(1)}\\
    &+ \sum_{\xi=2}^{2K-2} \sum_{k=1}^{N} \sum_{n_1,\cdots, n_k=2}^{\xi} 
	\updelta\left[\xi, \sum_{j=1}^k n_j\right]
	\delta q^{(k)} (\vec{n}),\\
	&= q^{(1)} + \sum_{\xi=2}^{2K-2} \sum_{k=1}^{K} \sum_{\vec{n} \in G_2(k, \xi)} \delta q^{(k)} (\vec{n}),
 \label{eq:setDefine}
\end{align}
where
\begin{equation}
\begin{split}	
\label{eq:qXiK}
	&\delta q^{(k)} (\vec{n}):=  
	\frac{\tau^{\sum_{j=1}^k n_j }}{n_1!n_2!\cdots n_k!} \\
	&~~~~\times {\rm Tr}\left(Q \mixture_{k,N-k}
        \left(
        \left(\mathcal{L}^{(n_1)},\ldots,\mathcal{L}^{(n_k)}\right)
        ,\mathcal{E}_N
        \right) (\rho_{\rm init})\right),
\end{split}
\end{equation}
with $\vec{n} = \{n_1, n_2, \cdots, n_k \}$ and where 
 we replace $N$ in the second summation with $K$ in the second equality (since the terms with $k > K$ become zero by the Kronecker $\updelta$.). To obtain \eqref{eq:setDefine}, we define the set $G_2(k, \xi)$ composed of all vectors with $k$ integer elements $\{n_j\}_{j=1}^k$ that satisfies $n_j \geq 2$ and $\sum_{j=1}^k n_j = \xi$. 
As an example, when using the second-order ($K=2$) qSWIFT, the summation of the second term in \eqref{eq:qXiK} includes only one term as:
\begin{equation}
	q^{(2)} = q^{(1)} + \delta q^{(1)}\left(\{2\}\right). 
\end{equation}
we see $\delta q$ in \eqref{eq:q2} corresponds to $\delta q^{(1)}\left(\{2\}\right)$, 
where we write $\{a_1, \cdots ,a_k\}$ as the vector having elements $a_1, \cdots ,a_k$.
As another example, when using the third-order ($K=3$) qSWIFT, 
\begin{equation}
\begin{split}	
	q^{(3)} &= q^{(1)} + \delta q^{(1)}\left(\{2\}\right) \\
	&~~~+ \delta q^{(1)}\left(\{3\}\right) + \delta q^{(1)}\left(\{4\}\right) + \delta q^{(2)}\left(\{2, 2\}\right). 
\end{split}
\end{equation}
Since $q^{(1)}$ is again evaluable by using the original qDRIFT, we focus on the evaluation of $\delta q^{(k)}(\vec{n})$ in the following. 

Similar to the second-order case,  
we will transform $\delta q^{(k)}(\vec{n})$ as a sum of the terms evaluable with quantum circuits.
The mixture function $\mixture_{k, N-k}(\cdot)$ included in \eqref{eq:qXiK} is written as the summation of $\binom{N}{k}$ terms as
\begin{equation}
\label{eq:qXiKN}
	\delta q^{(k)} (\vec{n})
	= \frac{\tau^{\sum_{j=1}^k n_j }}{n_1!n_2!\cdots n_k!} 
	\sum_{\sigma \in S_{N,k}^{\rm sub}} \delta q_{\sigma}^{(k)}(\vec{n}), 
\end{equation}
where 
\begin{equation}
\label{eq:qKSigmaDef}
\begin{split}	
&\delta q_{\sigma}^{(k)}(\vec{n}) \\
&:= 
		{\rm Tr}\left(Q  \sortingDefault\left(
        \left(\mathcal{L}^{(n_1)},\ldots,\mathcal{L}^{(n_k)}\right)
        ,\mathcal{E}_N
        \right) (\rho_{\rm init})\right), 
\end{split}
\end{equation}
with $\sortingDefault$ as the sorting function defined in \eqref{eq:sorting-function}.

Now we move on to the calculation of $\delta q_{\sigma}^{(k)}
        \left(\vec{n}\right)$. To this end, let 
 \begin{align}
 	\mathcal{D}^{(n)}_{0} := \mathcal{L}^n,~\mathcal{D}^{(n)}_{1} := \sum_\ell p_\ell \mathcal{L}_\ell^n.
 \end{align}
We can write $\mathcal{L}^{(n)}$ as a linear combination of $\mathcal{D}_s^{(n)}$: 
 \begin{equation}
 \label{eq:d-expand}
 \mathcal{L}^{(n)} = 	\sum_{s=0}^1 	(-1)^{s}\mathcal{D}^{(n)}_{s}
 \end{equation}
as per \eqref{eq:Deltak}. 
Let two probability distributions:
\begin{align}
 \label{eq:prob-expand}
	P_0^{(n)} (\vec{\ell}) &= P_\textsc{mdrift}(\vec{\ell}, n),\\
	P_1^{(n)} (\vec{\ell}) &= 
	\begin{cases}
			p_\ell & \ell_1 = \cdots =\ell_n = \ell,\\
			0    & \ell_{a} \neq \ell_b ~\text{for}~\exists (a, b)
	\end{cases}, 
\end{align}
where $n$ specifies the number of vectors in $\vec{\ell}$ and we write the $a$th element of $\vec{\ell}$ as $\ell_a$.
We see that for $n=2$, \eqref{eq:prob-expand} is consistent with \eqref{eq:prob-expand-2}.
 Then it holds 
\begin{equation}
\label{eq:lNExpand}
	\mathcal{D}_{s}^{(n)} = \sum_{\vec{\ell}} P_{s}^{(n)}(\vec{\ell}) \mathcal{L}_n(\vec{\ell}), 
\end{equation}
with 
\begin{equation}
	\mathcal{L}_n(\vec{\ell}) := \mathcal{L}_{\ell_n} \cdots \mathcal{L}_{\ell_1},  
\end{equation}
which is consistent with \eqref{eq:twoL} for $n=2$.
By using the bilinearity of the sorting function, we obtain
\begin{equation}
\begin{split}
&\sortingDefault\left(
        \left(\mathcal{L}^{(n_1)},\ldots,\mathcal{L}^{(n_k)}\right)
        ,\mathcal{E}_N
        \right) \\
&= \sum_{s_1, \cdots ,s_k=0}^1 (-1)^{\sum_{c} s_c}
 \sortingDefault \left(\left(\mathcal{D}^{(n_1)}_{s_1}, \cdots ,\mathcal{D}^{(n_k)}_{s_k}\right), \mathcal{E}_N \right), \\
&= \sum_{s_1, \cdots ,s_k=0}^1  (-1)^{\sum_{c} s_c} \sum_{\vec{\ell_1}, \cdots \vec{\ell_k}}
P_{s_1}^{(n_1)}(\vec{\ell}_1) \cdots P_{s_k}^{(n_k)}(\vec{\ell}_k) \\
 &~~~~~~\sortingDefault \left(\left(\mathcal{L}_{n_1}(\vec{\ell}_1), \cdots ,\mathcal{L}_{n_k}(\vec{\ell}_k)\right), \mathcal{E}_N \right), 
\end{split}	
\end{equation}
 where we use \eqref{eq:d-expand} in the first equality and \eqref{eq:lNExpand} in the second equality. 
Substituting above into \eqref{eq:qKSigmaDef}, we obtain
\begin{equation}
\begin{split}	
\label{eq:qSigmaKExpansion}
\delta q_{\sigma}^{(k)}
        \left(\vec{n}\right) &= \sum_{s_1, \cdots ,s_k=0}^1  (-1)^{\sum_{c} s_c} \sum_{\vec{\ell_1}, \cdots \vec{\ell_k}} \\
&~~~P_{s_1}^{(n_1)}(\vec{\ell}_1) \cdots P_{s_k}^{(n_k)}(\vec{\ell}_k)
\delta q_{\sigma}^{(k)}\left(\vec{n}, \left(\vec{\ell_1}, \cdots, \vec{\ell_k}\right)\right), 
\end{split}
\end{equation}
where 
\begin{equation}
\begin{split}	
\label{eq:qkSigmaNDef}
&\delta q_{\sigma}^{(k)}\left(\vec{n}, \left(\vec{\ell_1}, \cdots, \vec{\ell_k}\right)\right) \\
&:=  {\rm Tr}\left(Q\sortingDefault \left(\left(\mathcal{L}_{n_1}(\vec{\ell}_1), \cdots ,\mathcal{L}_{n_k}(\vec{\ell}_k)\right), \mathcal{E}_N \right)(\rho_{\rm init})\right).
\end{split}
\end{equation}

As in Section~\ref{section:secondOrderImpl}, we compute the right hand side of \eqref{eq:qkSigmaNDef} by using the system with one ancilla qubit. Let
\begin{equation}
\label{eq:multiL}
	\tilde{\mathcal{L}}_n (\vec{\ell}) := \tilde{\mathcal{L}}_{\ell_n}\cdots \tilde{\mathcal{L}}_{\ell_1}.
\end{equation}
By repeatedly operating \eqref{eq:tilde-lj} with $j = \ell_1, \cdots, \ell_n$, we obtain 
 the operation of $\tilde{\mathcal{L}}_n (\vec{\ell})$ as 
\begin{equation}
\left(
	\begin{array}{cc}
		\cdot & \rho \\
		\rho & \cdot
	\end{array}
\right) \xmapsto{\tilde{\mathcal{L}}_n(\vec{\ell})}
\left(
	\begin{array}{cc}
		\cdot & \mathcal{L}_n(\vec{\ell})(\rho) \\
		\mathcal{L}_n(\vec{\ell})(\rho) & \cdot
	\end{array}
\right)  
\end{equation}
for a given density operator $\rho$. 
Recall the definition of the sorting function \eqref{eq:sorting-function}:
\begin{equation}
\begin{split}	
	&\sortingDefault \left(\left(\mathcal{L}_{n_1}(\vec{\ell}_1), \cdots ,\mathcal{L}_{n_k}(\vec{\ell}_k)\right), \mathcal{E}_N \right) \\
	&=  
	\mathcal{X}_{\sigma(1)} \cdots \mathcal{X}_{\sigma(N)}, 
\end{split}
\end{equation}
with 
\begin{equation}
\mathcal{X}_a = 
\begin{cases}
 \mathcal{L}_{n_a}(\vec{\ell}_a) & a\leq k,\\
 \mathcal{E}_N & a\geq k+1
\end{cases}.
\end{equation}
Then in the system with one ancilla qubit, it holds 
\begin{equation}
\begin{split}	
	&\sortingDefault \left(\left(\tilde{\mathcal{L}}_{n_1}(\vec{\ell}_1), \cdots ,\tilde{\mathcal{L}}_{n_k}(\vec{\ell}_k)\right), \tilde{\mathcal{E}}_N \right) \\
	&= 
	\tilde{\mathcal{X}}_{\sigma(1)} \cdots \tilde{\mathcal{X}}_{\sigma(N)}, 
\end{split}
\end{equation}
with 
\begin{equation}
\tilde{\mathcal{X}}_a = 
\begin{cases}
 \tilde{\mathcal{L}}_{n_a}(\vec{\ell}_a) & a\leq k,\\
 \tilde{\mathcal{E}}_N & a\geq k+1
\end{cases}.
\end{equation}
Since it holds that 
\begin{equation}
	\left(
	\begin{array}{cc}
		\cdot & \rho \\
		\rho & \cdot
	\end{array}
\right) \xmapsto{\tilde{\mathcal{X}}_a}
\left(
	\begin{array}{cc}
		\cdot & \mathcal{X}_a(\rho) \\
		\mathcal{X}_a(\rho) & \cdot
	\end{array}
\right),   
\end{equation}
the operation of $\tilde{\mathcal{X}}_{\sigma(1)} \cdots \tilde{\mathcal{X}}_{\sigma(N)}$ to the input state $\tilde{\rho}_{\rm init} = |+\rangle\langle +| \otimes \rho_{\rm init}$ reproduces the 
operation of $\mathcal{X}_{\sigma(1)} \cdots \mathcal{X}_{\sigma(N)}$ as 
\begin{equation}
\begin{split}	
    \label{eq:higherOrderTransform}
	&\tilde{\rho}_{\rm init} = 
	\left(
	\begin{array}{cc}
		\rho_{\rm init}/2 & \rho_{\rm init}/2 \\
		\rho_{\rm init}/2 & \rho_{\rm init}/2 
	\end{array}
\right) \xmapsto{\tilde{\mathcal{X}}_{\sigma(1)} \cdots \tilde{\mathcal{X}}_{\sigma(N)}} \\
 &
 	~~~~~\left(
	\begin{array}{cc}
		\cdot & \mathcal{X}_{\sigma(1)} \cdots \mathcal{X}_{\sigma(N)}(\rho_{\rm init})/2 \\
	\mathcal{X}_{\sigma(1)} \cdots \mathcal{X}_{\sigma(N)}(\rho_{\rm init})/2 & \cdot
	\end{array}
\right). 
\end{split}
\end{equation}
Therefore, the estimation value of the observable $\tilde{Q} = X \otimes Q$ with the final state in \eqref{eq:higherOrderTransform} gives $\delta q_{\sigma}^{(k)}\left(\vec{n}, \left(\vec{\ell_1}, \cdots, \vec{\ell_k}\right)\right)$, i.e., 
 \begin{equation}
 \begin{split} 	
 \label{eq:delqOneAncilla}
&\delta q_{\sigma}^{(k)}\left(\vec{n}, \left(\vec{\ell_1}, \cdots, \vec{\ell_k}\right)\right) \\
 	&= 
 	 	{\rm Tr}\left(\tilde{Q} \sortingDefault \left(\left(\tilde{\mathcal{L}}_{n_1}(\vec{\ell}_1), \cdots ,\tilde{\mathcal{L}}_{n_k}(\vec{\ell}_k)\right), \tilde{\mathcal{E}}_N \right)(\tilde{\rho}_{\rm init}) \right). 
 \end{split}	
 \end{equation}
 
Next, we discuss the way of evaluating the right hand side of \eqref{eq:delqOneAncilla}. 
By substituting \eqref{eq:l-sum} to \eqref{eq:delqOneAncilla}, we obtain
\begin{equation}
\label{eq:lTildeExpand}
	\tilde{\mathcal{L}}_n(\vec{\ell}) = \sum_{\vec{b}} \tilde{\mathcal{S}}_{n}^{(\vec{b})} (\vec{\ell}),
\end{equation}
with $\vec{b} \in \{0, 1\}^{\otimes n}$, 
where 
 \begin{equation}
 	\tilde{\mathcal{S}}_{n}^{(\vec{b})} (\vec{\ell}) = \tilde{\mathcal{S}}_{\ell_n}^{(b_n)} \cdots \tilde{\mathcal{S}}_{\ell_1}^{(b_1)}. 
 \end{equation}
Then with $\vec{b}_j \in \{0, 1\}^{\otimes n_j} (j = 1\cdots k)$, we obtain
\begin{equation}
	 \label{eq:higherUnitarySum}
\begin{split}	
	 &\sortingDefault \left(\left(\tilde{\mathcal{L}}_{n_1}(\vec{\ell}_1), \cdots ,\tilde{\mathcal{L}}_{n_k}(\vec{\ell}_k)\right), \tilde{\mathcal{E}}_N \right) \\
	 &= \sum_{\vec{b_1}\cdots \vec{b_k}}\sortingDefault \left(\left(\tilde{\mathcal{S}}_{n_1}^{(\vec{b_1})}(\vec{\ell}_1), \cdots ,\tilde{\mathcal{S}}_{n_k}^{(\vec{b_k})}(\vec{\ell}_k)\right), \tilde{\mathcal{E}}_N \right) \\
	 &= \sum_{\vec{b_1}\cdots \vec{b_k}} \sum_{\vec{r}}
	 P_{\textsc{mdrift}}(\vec{r}, N-k) \\
	 &~~~~~~~~~\mathcal{C}_{\sigma,k, N-k}^{(\vec{b}_1, \cdots ,\vec{b}_k)}
	 \left(\vec{n}, \left(
	 \vec{\ell}_1, \cdots \vec{\ell}_k
	 \right), \vec{r}\right),
\end{split}
\end{equation}
where $\mathcal{C}_{\sigma,k, N-k}^{(\vec{b}_1, \cdots ,\vec{b}_k)}
	 \left(\vec{n}, \left(
	 \vec{\ell}_1, \cdots \vec{\ell}_k
	 \right), \vec{r}\right)$ is an unitary channel defined by 
\begin{equation}
\hspace{-1cm}
\begin{split}	
 &\mathcal{C}_{\sigma,k, N-k}^{(\vec{b}_1, \cdots ,\vec{b}_k)}
	 \left(\vec{n}, \left(
	 \vec{\ell}_1, \cdots \vec{\ell}_k
	 \right), \vec{r}\right) \\
&= \sortingDefault \left(\left(\tilde{\mathcal{S}}_{n_1}^{(\vec{b_1})}(\vec{\ell}_1), \cdots ,\tilde{\mathcal{S}}_{n_k}^{(\vec{b_k})}(\vec{\ell}_k)\right), \left(\tilde{\mathcal{T}}_{r_1}, \cdots, \tilde{\mathcal{T}}_{r_{N-k}} \right) \right). 
\end{split}
\end{equation}
We use \eqref{eq:lTildeExpand}  in the first equality of \eqref{eq:higherUnitarySum} and we use 
\begin{equation}
	\tilde{\mathcal{E}}_N := \sum_{r=1}^L p_{r} \tilde{\mathcal{T}}_r, 
\end{equation} 
and 
\begin{equation}
	P_{\textsc{mdrift}}(\vec{r}, N-k) = p_{r_1} \cdots p_{r_{N-k}},
\end{equation}
with $\vec{\ell} = \{\ell_1 \cdots \ell_{N-k}\}$
in the second equality.
By substituting \eqref{eq:higherUnitarySum} to 
\eqref{eq:delqOneAncilla}, we obtain
\begin{equation}
\label{eq:qSigmaNJ}
\begin{split}	
&\delta q_{\sigma}^{(k)}\left(\vec{n}, \left(\vec{\ell_1}, \cdots, \vec{\ell_k}\right)\right) \\
&= \sum_{\vec{b_1}\cdots \vec{b_k}}\sum_{\vec{r}} 
P_{\textsc{mdrift}}(\vec{r}, N-k) \\
&~~~~{\rm Tr}\left(\tilde{Q} \mathcal{C}_{\sigma,k, N-k}^{(\vec{b}_1, \cdots ,\vec{b}_k)}
	 \left(\vec{n}, \left(
	 \vec{\ell}_1, \cdots \vec{\ell}_k
	 \right), \vec{r}\right)
	  (\tilde{\rho}_{\rm init}) \right).
\end{split}
\end{equation}
Finally, combining \eqref{eq:qXiKN}, \eqref{eq:qKSigmaDef}, and \eqref{eq:qSigmaKExpansion} with \eqref{eq:qSigmaNJ}, 
\begin{equation}
	\begin{split}		
\label{eq:qXiKFinal}
	&\delta q^{(k)} (\vec{n}) \\
	&= \binom{N}{k}\frac{\tau^{\sum_{j=1}^k n_j }}{n_1!n_2!\cdots n_k!} 
	\frac{1}{\binom{N}{k}} \sum_{\sigma \in S_{N,k}^{\rm sub}} \delta q_{\sigma}^{(k)}(\vec{n}), \\
	&= c^{(k)}(\vec{n})\sum_{\vec{b_1}\cdots \vec{b_k}}  \sum_{\vec{s}} (-1)^{\sum_{c=1}^k s_c} \delta q^{(k)}\left(\left(\vec{b_1}, \cdots \vec{b_k}\right), \vec{s} \right), 
	\end{split}
\end{equation}
where 
\begin{equation}
\label{eq:qSB}
\begin{split}	
	&\delta q^{(k)}\left(\left(\vec{b_1}, \cdots \vec{b_k}\right), \vec{s} \right) \\
	&:= \frac{1}{\binom{N}{k}}
	\sum_{\sigma \in S_{N,k}^{\rm sub}}  
	\sum_{\vec{\ell_1}, \cdots \vec{\ell_k}}
P_{s_1}^{(n_1)}(\vec{\ell}_1) \cdots P_{s_k}^{(n_k)}(\vec{\ell}_k) \\
	&~~~~\sum_{\vec{r}}P_{\textsc{mdrift}}(\vec{r}, N-k) \\
	&~~~~{\rm Tr}
	 \left(
	 \tilde{Q}
	 \mathcal{C}_{\sigma,k, N-k}^{(\vec{b}_1, \cdots ,\vec{b}_k)}
	 \left(\vec{n}, \left(
	 \vec{\ell}_1, \cdots \vec{\ell}_k
	 \right), \vec{r}\right)
	 (\tilde{\rho}_{\rm init}) 
	 \right), 
\end{split}
\end{equation}
and we define the coefficient as
\begin{equation}
	c^{(k)}(\vec{n}) :=  \binom{N}{k}\frac{\tau^{\sum_{j=1}^k n_j }}{n_1!n_2!\cdots n_k!}. 
\end{equation}

\begin{figure*}
\begin{minipage}{\linewidth}
 \begin{algorithm}[H]
  \caption{{\bf Evalcorrection}}
  \label{algorithm:high-order-1}
  \begin{algorithmic}[1]
  	\INPUT  
  	$k$, $\vec{n}$, $N$, $N_{\rm sample}$, and $N_{\rm shot}$
  	 \FOR{$\vec{s}$ in $\{0, 1\}^{\otimes k}$ }
  	 \FOR{($\vec{b}_1, ..., \vec{b}_k)$ in $(\{0, 1\}^{\otimes n_1}, ..., \{0, 1\}^{\otimes n_k})$}
  	   	\FOR{$p = 1$ to $N_{\rm sample}$}
  	   	  	 \STATE Sample $\sigma$ uniformly from $S_{N,k}^{\rm sub}$.
  	         \STATE Sample $\vec{\ell}_1, \cdots \vec{\ell}_k$ according to $P_{s_1}^{n_1}(\vec{\ell}_1) \cdots P_{s_k}^{n_k}(\vec{\ell}_k)$. Sample $\vec{r}$ according to $P_{\textsc{mdrift}}(\vec{r}, N-k)$.
	         \STATE With $N_{\rm shot}$ measurements, estimate the following by the quantum circuit and set it to $\delta \hat{q}_{\sigma_p}^{(k)}\left(\left(\vec{b_1}, \cdots \vec{b_k}\right), \vec{s} \right)$: 
	\begin{equation}
		{\rm Tr}
	 \left(
	 \tilde{Q}
	 \mathcal{C}_{\sigma,k, N-k}^{(\vec{b}_1, \cdots ,\vec{b}_k)}
	 \left(\vec{n}, \left(
	 \vec{\ell}_1, \cdots \vec{\ell}_k
	 \right), \vec{r}\right)
	 (\tilde{\rho}_{\rm init}) 
	 \right), 
		\nonumber
	\end{equation}
	\ENDFOR
	\STATE Set $\delta \hat{q}^{(k)}\left(\left(\vec{b_1}, \cdots \vec{b_k}\right), \vec{s} \right) =  \frac{1}{N_{\rm sample}}\sum_{p=1}^{N_{\rm sample}}\delta \hat{q}_{\sigma_p}^{(k)}\left(\left(\vec{b_1}, \cdots \vec{b_k}\right), \vec{s} \right)$.
	\ENDFOR
  	\ENDFOR
  	\STATE Set 
  	\begin{equation}
  	\nonumber
  		   	\delta \hat{q}^{(k)} (\vec{n})= 
c^{(k)}(\vec{n}) \sum_{\vec{b_1}\cdots \vec{b_k}}  \sum_{\vec{s}} (-1)^{\sum_{c=1}^k s_c}
 \delta \hat{q}^{(k)}\left(\left(\vec{b_1}, \cdots \vec{b_k}\right), \vec{s} \right).
  	\end{equation}
  	\OUTPUT $	\delta \hat{q}^{(k)} (\vec{n})$
  \end{algorithmic}
\end{algorithm}
\end{minipage}
\end{figure*}

\begin{figure*}
\begin{minipage}{\linewidth}
 \begin{algorithm}[H]
  \caption{{\bf qSWIFT}}
  \label{algorithm:qSWIFT}
  \begin{algorithmic}[1]
  	\INPUT $K$, $N$, $N_{\rm sample}(\vec{n})$, $N_{\rm shot}(\vec{n})$, $N_{\rm sample}^{0}$, and $N_{\rm shot}^{0}$
  	\STATE Estimate ${\rm Tr}(Q\mathcal{E}_N^N(\rho_{\rm init}))$ by sampling $N_{\rm sample}^{0}$ circuits and measuring each circuit $N_{\rm shot}^{0}$ times; set the result to $\hat{q}^{(1)}$.
  	\STATE Set delta = 0.
  	\FOR{$\xi = 2,...,2K-2$}
  	\FOR{$k = 1,...,K$}
  	\FOR{$\vec{n} \in G_2(k, \xi)$}
  	\STATE Add the result of ${\bf Evalcorrection}\left(k, \vec{n}, N, N_{\rm sample}(\vec{n}), N_{\rm shot}(\vec{n})\right)$ to delta.
  	\ENDFOR
  	\ENDFOR
  	\ENDFOR
  	\STATE Set $\hat{q}^{(K)} = \hat{q}^{(1)} + {\rm delta}$.
  	\OUTPUT $\hat{q}^{(K)}$
  \end{algorithmic}
\end{algorithm}
\end{minipage}
\end{figure*}

We can get an unbiased estimator of \eqref{eq:qSB} using Monte Carlo sampling. 
More specifically, we repeat the following procedure and compute the average of the output; the $p$-th operation works as follows:
\begin{enumerate}
	\item Sample $\sigma$ uniformly from all elements of $S_{N,k}^{\rm sub}$. 
	\item Sample $\vec{\ell}_1, \cdots \vec{\ell}_k$ according to $P_{s_1}^{n_1}(\vec{\ell}_1) \cdots P_{s_k}^{n_k}(\vec{\ell}_k)$. Sample $\vec{r}$ according to $P_{\textsc{mdrift}}(\vec{r}, N-k)$.
	\item Evaluate 
	\begin{equation}
		{\rm Tr}
	 \left(
	 \tilde{Q}
	 \mathcal{C}_{\sigma,k, N-k}^{(\vec{b}_1, \cdots ,\vec{b}_k)}
	 \left(\vec{n}, \left(
	 \vec{\ell}_1, \cdots \vec{\ell}_k
	 \right), \vec{r}\right)
	 (\tilde{\rho}_{\rm init}) 
	 \right), 
	\end{equation}
	with $N_{\rm shot}$ measurements. We set the result to $\delta \hat{q}_{\sigma_p}^{(k)}\left(\left(\vec{b_1}, \cdots \vec{b_k}\right), \vec{s} \right)$.
\end{enumerate}
As in the second-order case, we repeat the above process $N_{\rm sample}$ times and compute the average of $\delta \hat{q}_{\sigma_p}^{(k)}\left(\left(\vec{b_1}, \cdots \vec{b_k}\right), \vec{s} \right)$ as $\delta \hat{q}^{(k)}\left(\left(\vec{b_1}, \cdots \vec{b_k}\right), \vec{s} \right)$, which gives the estimate of $\delta q^{(k)}\left(\left(\vec{b_1}, \cdots \vec{b_k}\right), \vec{s} \right)$. 
Substituting the estimate to \eqref{eq:qXiKFinal}, we obtain the estimate of $\delta q^{(k)} (\vec{n})$ as 
\begin{equation}
\begin{split}	
&\delta \hat{q}^{(k)} (\vec{n}) \\
&= 
c^{(k)}(\vec{n}) \sum_{\vec{b_1}\cdots \vec{b_k}}  \sum_{\vec{s}} (-1)^{\sum_{c=1}^k s_c}
 \delta \hat{q}^{(k)}\left(\left(\vec{b_1}, \cdots \vec{b_k}\right), \vec{s} \right).
\end{split}
\end{equation}
We write the above algorithm to estimate $\delta q^{(k)} (\vec{n})$ as {\bf Evalcorrection} and 
summarize it in {\bf Algorithm \ref{algorithm:high-order-1}}.
By using the {\bf Evalcorrection}, we can construct the algorithm to compute $q^{(K)}$ according to \eqref{eq:allOrder}. We write the algorithm as {\bf qSWIFT} and summarize it in {\bf Algorithm~\ref{algorithm:qSWIFT}}. Since we can tune $N_{\rm sample}$ and $N_{\rm shot}$ depending on $\delta \hat{q}^{(k)}(\vec{n})$ we parameterize them as $N_{\rm sample}(\vec{n})$ and $N_{\rm shot}(\vec{n})$. We also write the number of circuits sampled and the number of running each circuit for calculating $q^{(1)}$ as $N_{\rm sample}^0$ and $N_{\rm shot}^0$.

\figthree{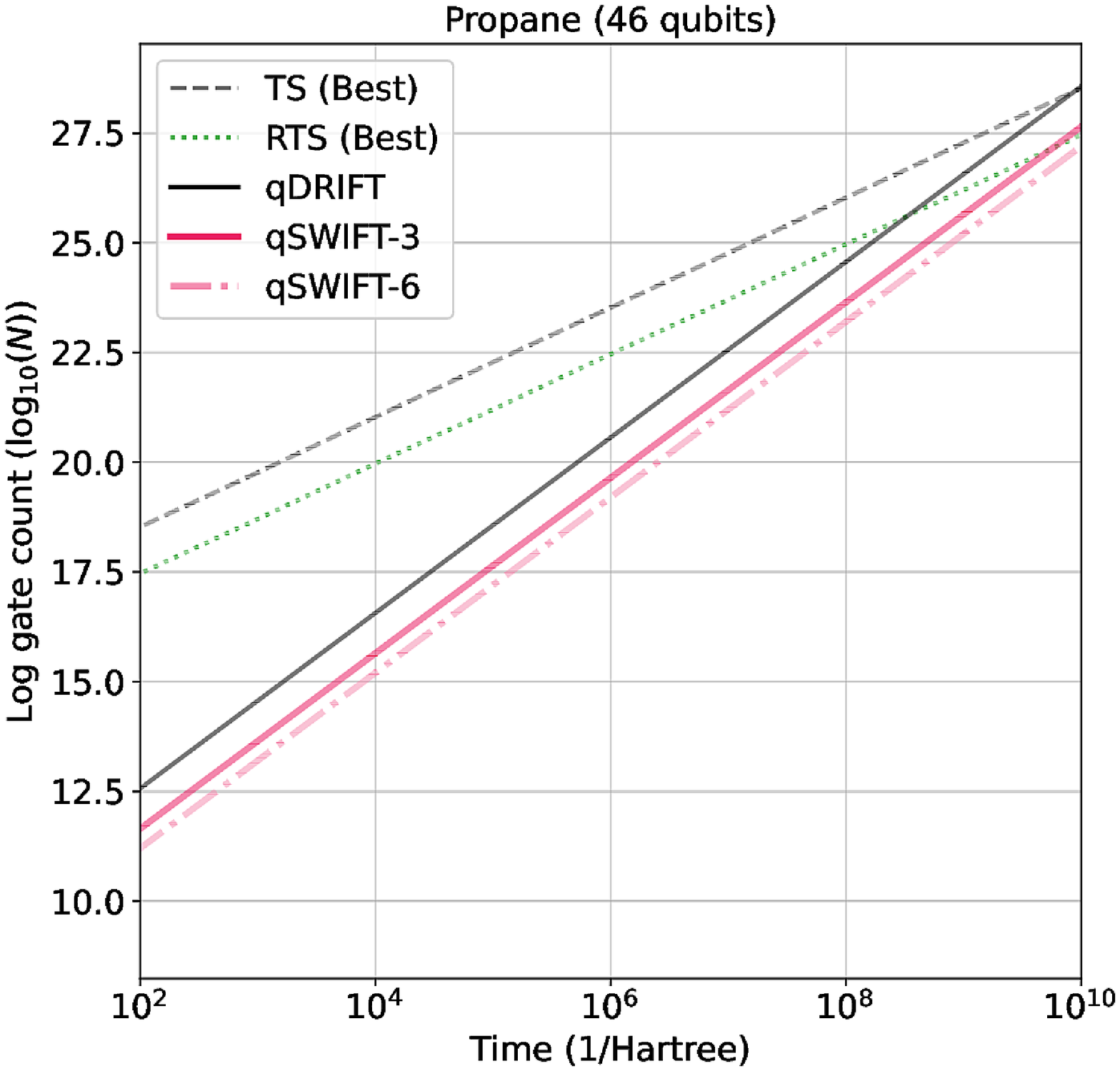}{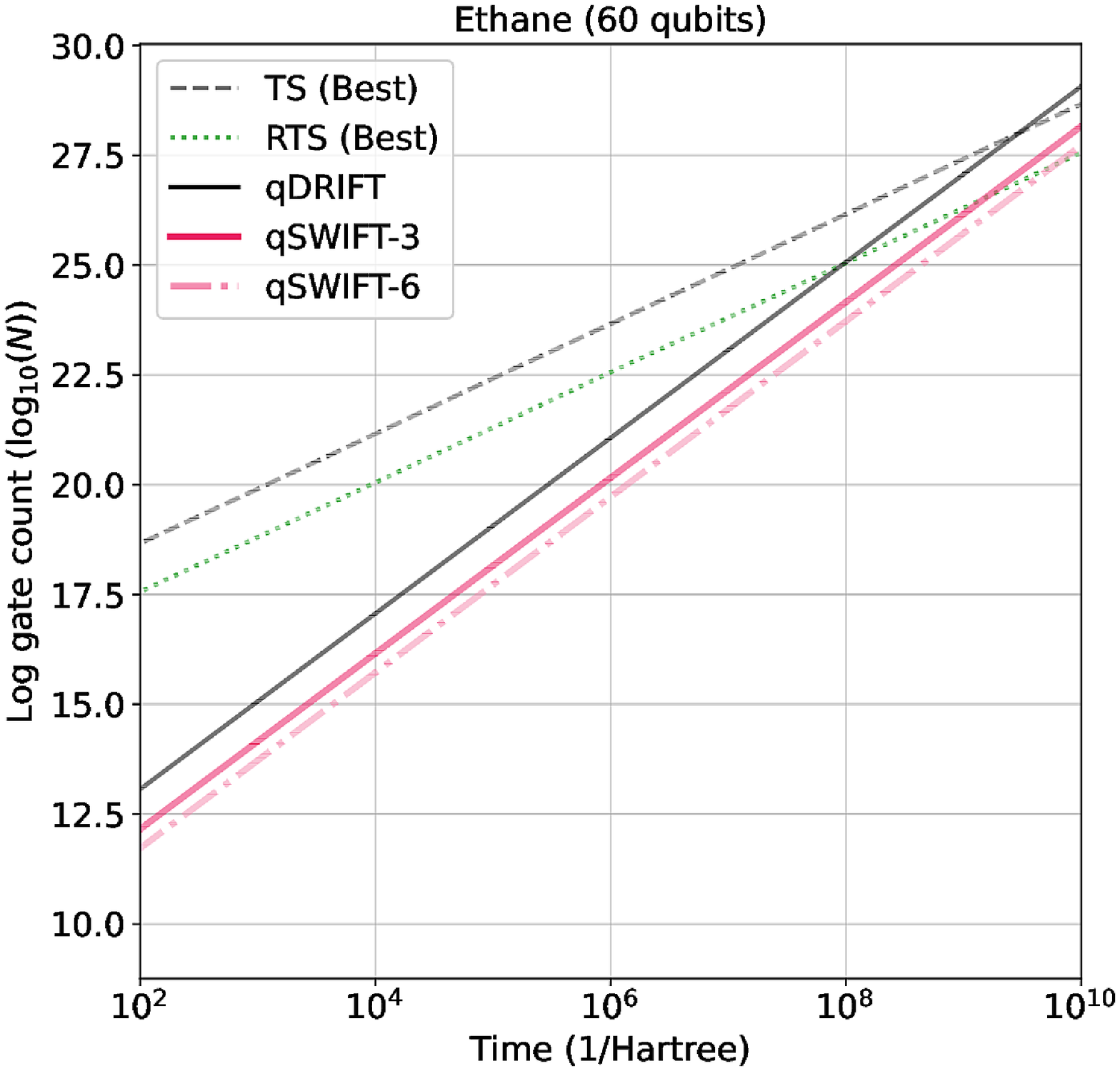}{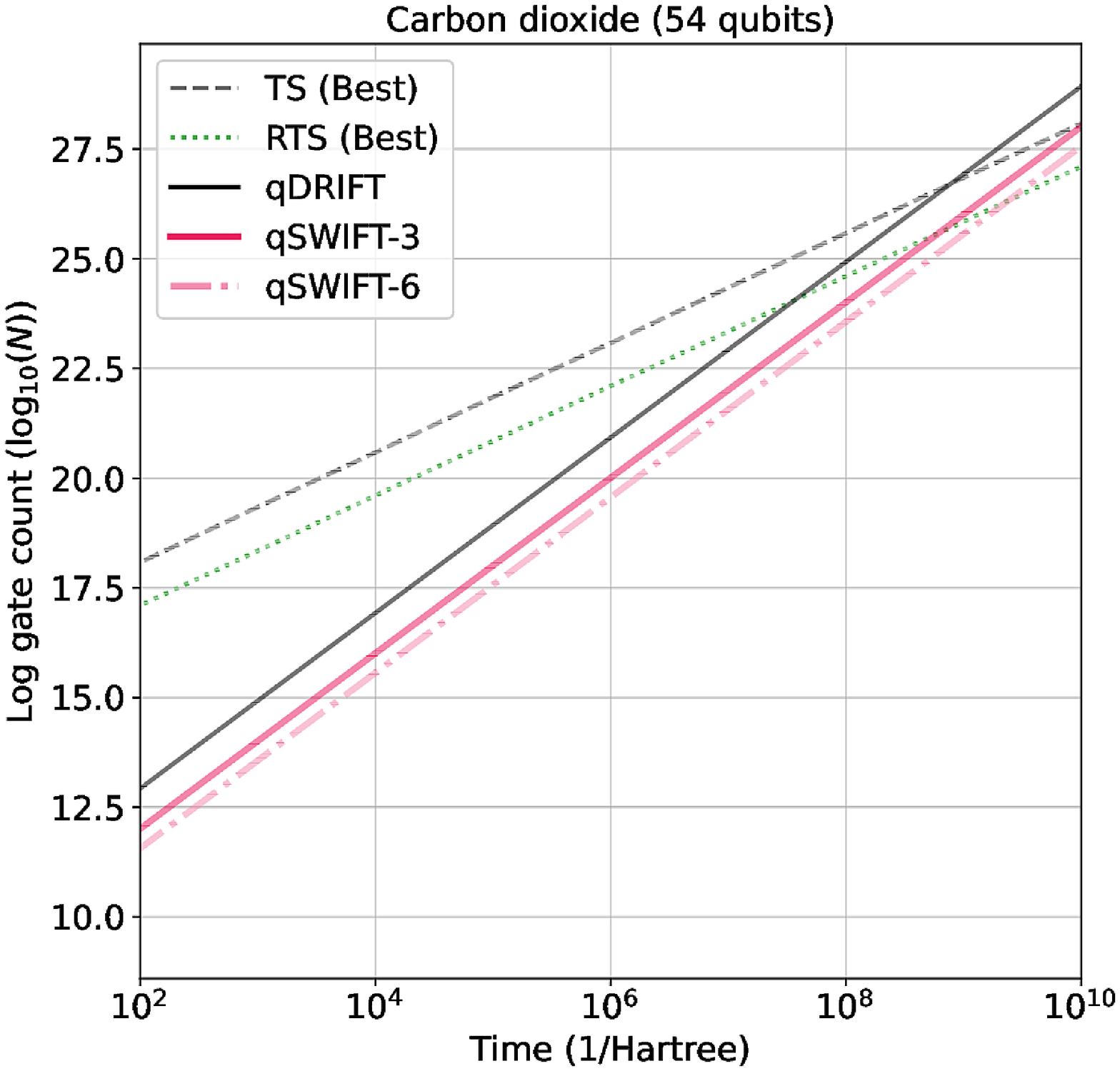}{160pt}{The asymptotic behaviors of $N$ for each time for each molecule to achieve the systematic error $\varepsilon = 0.001$. Three subfigures correspond to each molecule: propane with STO-3G basis (left), ethane with 6-31G basis (center), and carbon dioxide with the 6-31G basis (right). We show the third-order qSWIFT by the pink line (qSWIFT-3), the sixth-order qSWIFT by the pink dotted line (qSWIFT-6), and the qDRIFT by the black line (qDRIFT). For the Trotter-Suzuki decomposition, the best of the deterministic method among the first, second, and fourth order is shown by the gray dotted line (TS (Best)), and the best of the randomized method is shown by the green dotted line (RTS (Best)). }{fig:asymptotic}

\figthree{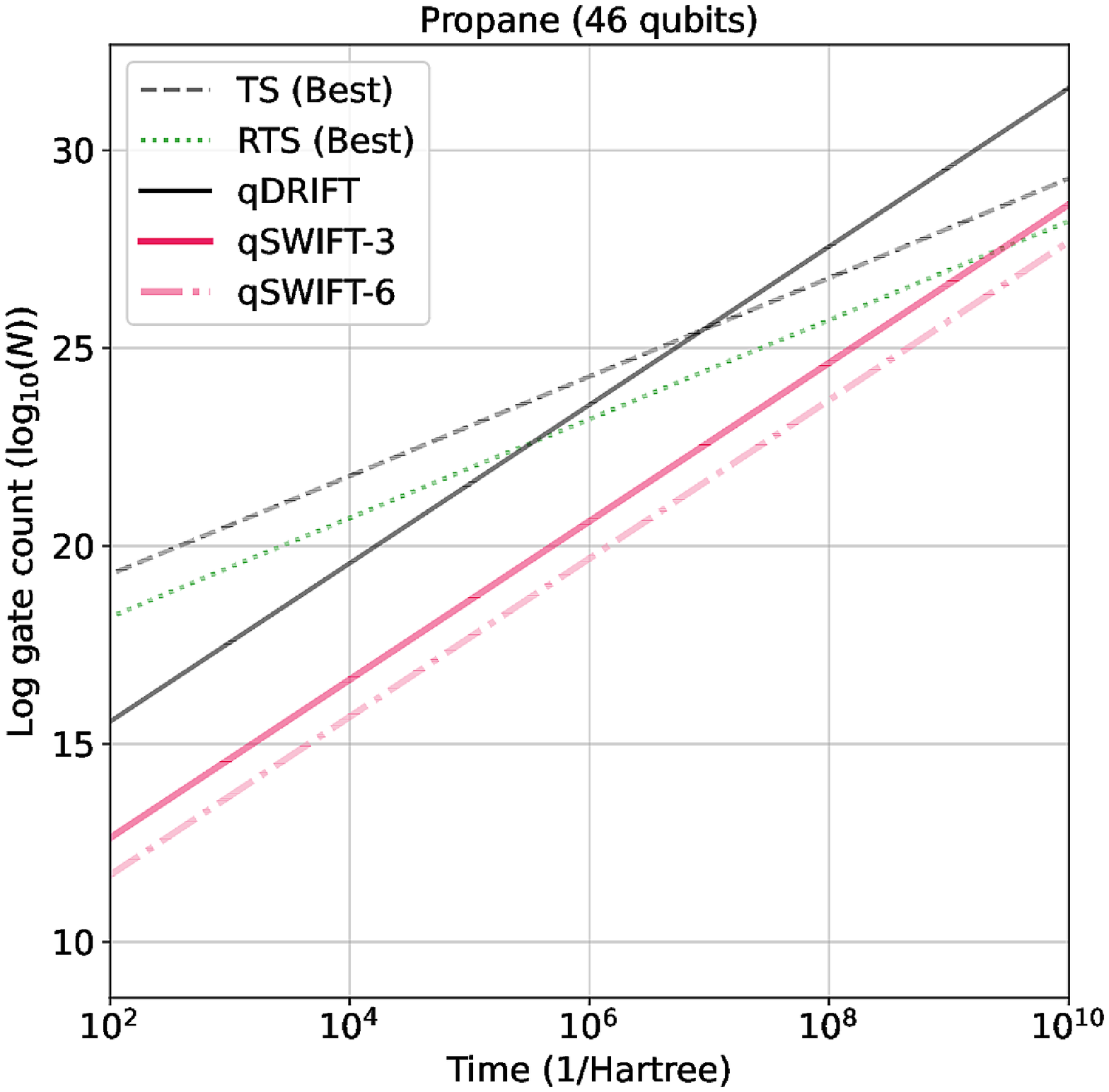}{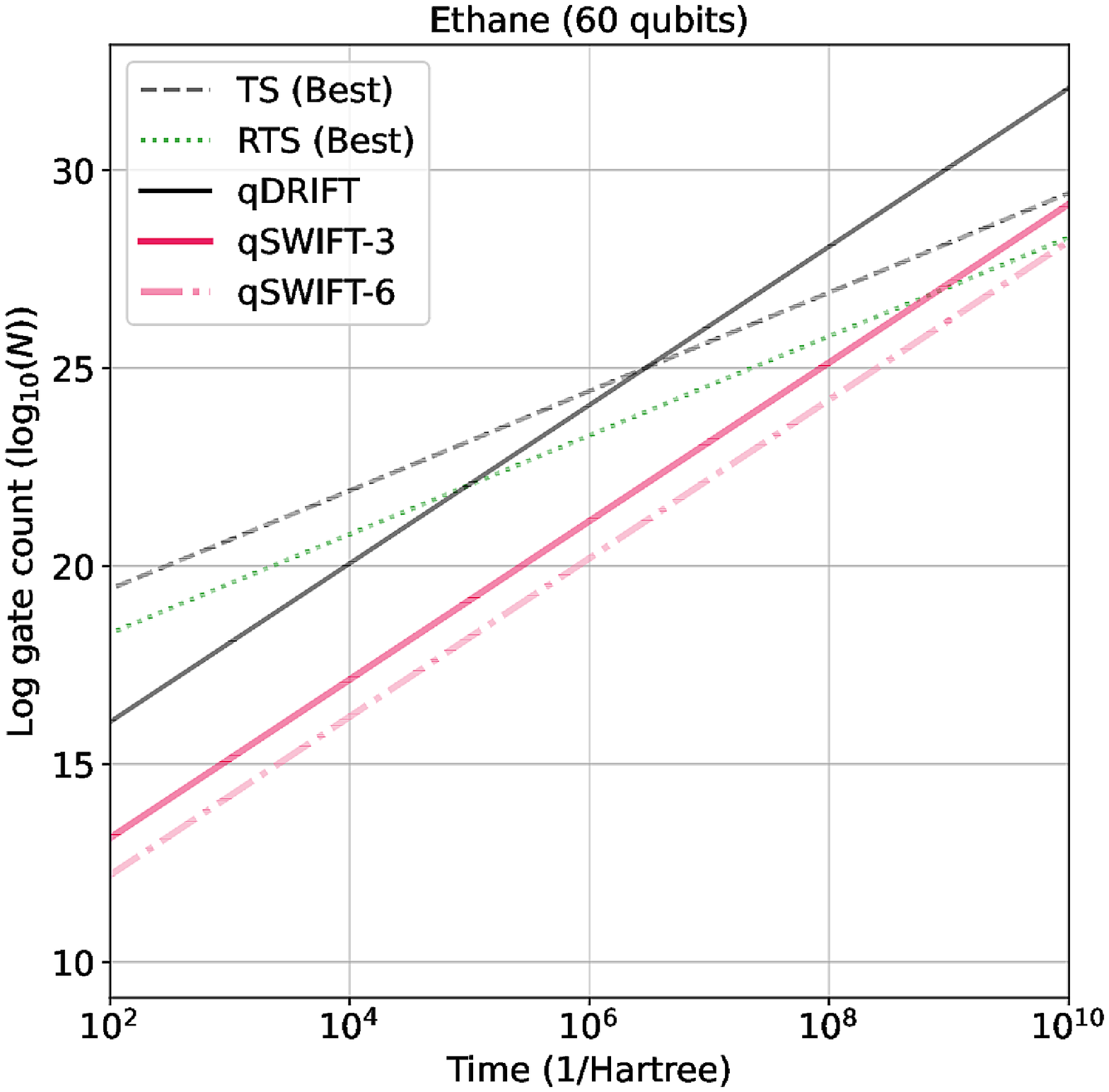}{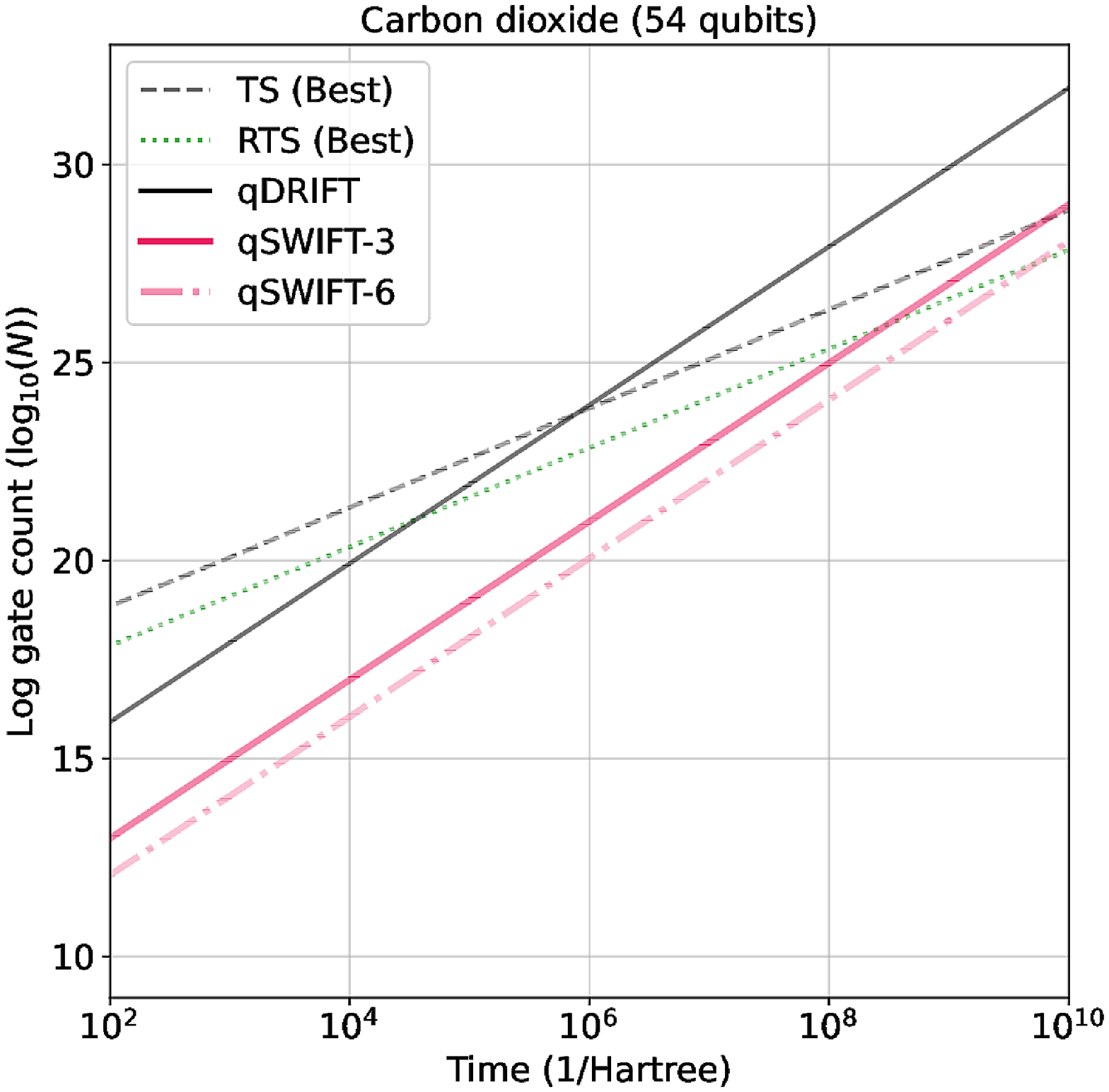}{160pt}{The asymptotic behavior to achieve $\varepsilon = 10^{-6}$. Other settings are the same as Fig.~\ref{fig:asymptotic}.}{fig:asymptoticPrecise}

It should be noted that there is a statistical error in $\hat{q}^{(K)}$ due to the use of Monte Carlo sampling and the limited number of measurements. To reduce the statistical error, we need to increase $N_{\rm sample}(\vec{n})$, $N_{\rm shot}(\vec{n})$, $N_{\rm sample}^{0}$, and $N_{\rm shot}^{0}$. We show how the total number of quantum circuits runs scale with the order in Appendix~\ref{section:statistical} even though our focus in this paper is discussing the reduction of systematic error and not the statistical error. In the discussion, we show that the dominant source of the quantum circuit runs comes from the estimation of $\Delta\hat{q}^{(k)}(\vec{n})$ with limited $\vec{n}$ and the cost for $\Delta\hat{q}^{(k)}(\vec{n})$ with other $\vec{n}$ is negligible when $N$ is large; consequently, the total number of quantum circuit runs scale only less than quadratically with $K$. 

\knadd{
In this section, we present the high-order qSWIFT method by expanding the unitary channel $\mathcal{U}$, where the systematic error decreases exponentially with the order parameter. We wish to highlight that it is possible to construct an all-order qSWIFT, which completely eliminates systematic error, as shown in Appendix~\ref{section:allOrderqSWIFT}. To avoid exponentially large statistical errors, we must set $N \in \mathcal{O}\left((\lambda t)^2\right)$. The primary drawback of the all-order qSWIFT is the lack of an upper bound for the number of swift operators. Therefore, the higher-order qSWIFT is more appropriate for use in quantum devices with limited gate operation capacity. The detail of the all-order qSWIFT is described in Appendix~\ref{section:allOrderqSWIFT}.

\subsection{Note on the previous higher-order randomized method}

We mention that the work by Wan et al.~\cite{wan2022randomized} introduces a higher-order randomized technique for phase estimation, where the task is to compute ${\rm Tr}[\rho e^{\upi Ht}]$ with~$\rho$ as the initial state.
They propose a randomized approach to estimate ${\rm Tr}\left[\rho e^{\upi Ht} \right]$ that leverages the linear combination of unitaries (LCU) scheme.
Methodologically, they express $e^{\upi Ht}$ as a weighted sum of unitary operations and select a term for sampling based on a specific probability distribution.

While the study \cite{wan2022randomized} is primarily focused on phase estimation, it could be extended to a higher-order randomization approach to estimate 
$q = {\rm Tr}(Q \mathcal{U}(\rho_{\rm init})) = {\rm Tr}(Q e^{\upi Ht}\rho_{\rm init} e^{-\upi Ht})$ by expanding $e^{\upi Ht}$ and $e^{-\upi Ht}$ in a similar fashion to the LCU technique, which we refer to as the LCU-based method. Detailed in Appendix~\ref{section:lcuBased}, this method, as in the all-order qSWIFT presented in Appendix~\ref{section:allOrderqSWIFT}, is free from systematic error. Nonetheless, the LCU-based method demands that all the $\mathcal{O}((\lambda t)^2)$ time-evolution operations to be controlled by the ancilla qubit.
In contrast, qSWIFT only requires swift operators to interact with the ancilla qubit.  Appendix~\ref{section:lcuBased} illustrates how the demand for $\mathcal{O}((\lambda t)^2)$ time evolutions in the LCU-based method could lead to a substantial increase in the number of control gates in comparison to qSWIFT for certain problems.
}

\section{Numerical experiments}
\label{section:numerics}
In this section, we present two numerical simulations of qSWIFT. 
In Section~\ref{section:asymptotic}, we show the asymptotic behavior of qSWIFT and compare it with Trotter-Suzuki decomposition and qDRIFT. In Section~\ref{section:simulation}, we perform a numerical experiment with the hydrogen molecule Hamiltonian and compare its performance with those of the previous algorithms.

\subsection{Asymptotic behaviour}
\label{section:asymptotic}

We compute the required number of gates $N$ to achieve a given systematic error for each evolution time and each algorithm: qSWIFT, Trotter-Suzuki decomposition, and qDRIFT. 
To clarify the difference from the original qDRIFT, we use the three molecules used in the numerical experiment of the original paper \cite{campbell2019random}: propane, ethane, and carbon dioxide.

For the qSWIFT algorithm, we utilize the third-order ($K=3$) qSWIFT. We also use the sixth-order qSWIFT  ($K=6$) as a reference. For the Trotter-Suzuki decomposition, we use both the deterministic and the randomized methods of the first, second, and fourth order. 
For calculating the systematic error of qSWIFT, we use the bound \eqref{eq:exactFinalBound} derived in Appendix~\ref{section:higherErrorBound}.
For the error of qDRIFT 
and the Trotter-Suzuki decompositions, we utilize the same upper-bound formulae as in 
the literature \cite{campbell2019random} (See Appendix~B and Appendix~C of \cite{campbell2019random}). 

\fig{250pt}{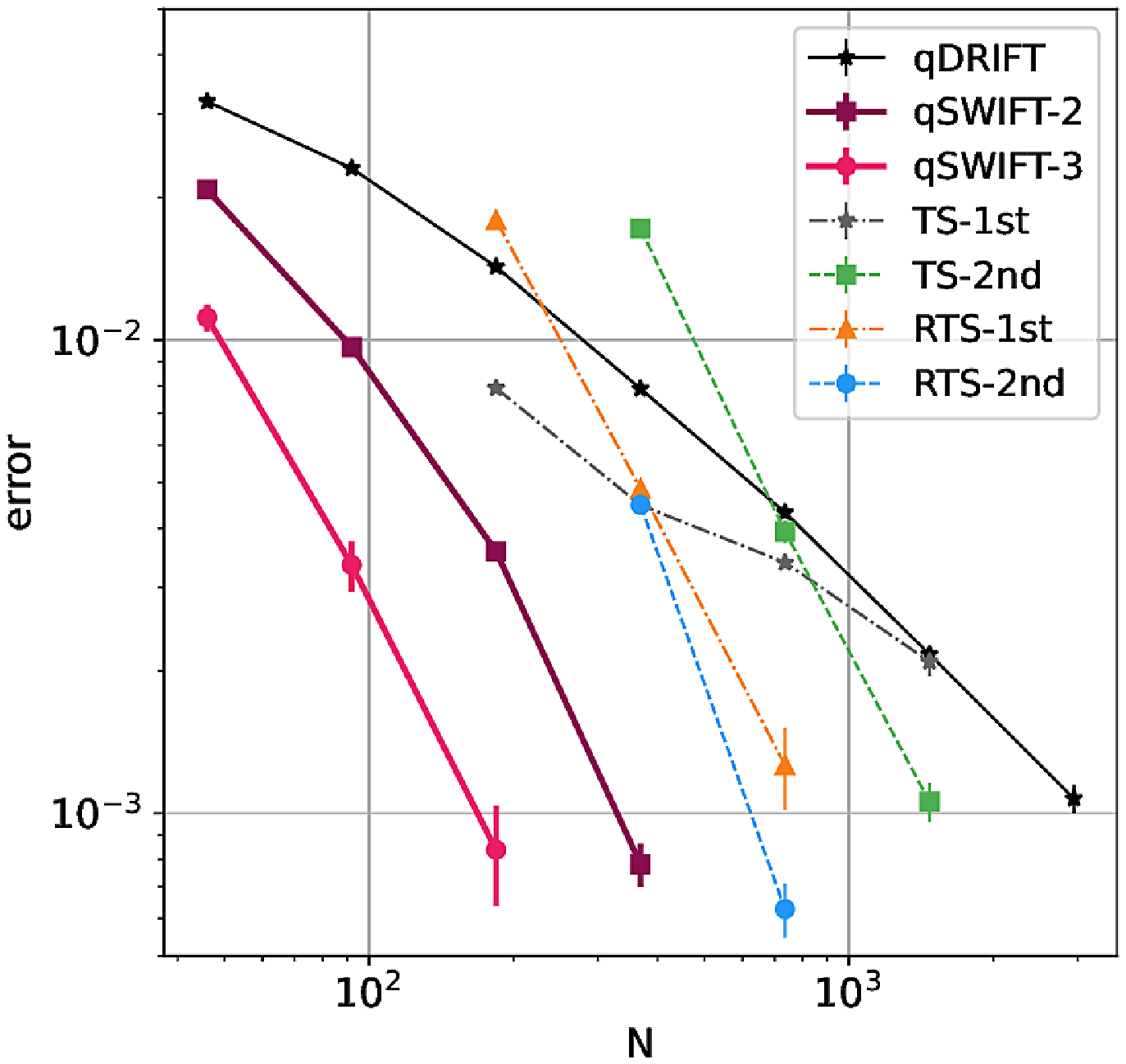}{The estimation error of ${\rm Tr}(Q\mathcal{U}(\rho_{\rm init}))$ for each number of gates $N$ for each method with $Q = ZIIIIIII$ and $\rho_{\rm init} = |+\rangle^{\otimes 8}$. The time evolution is performed by the hydrogen molecule Hamiltonian with 6-31g basis transformed by the Bravyi-Kitaev transformation. The evolution time is set to be $t=1$. In plotting each point, we perform six trials and show the mean value and the standard deviation of the mean. For qSWIFT, we show the result of the second-order (qSWIFT-2) with the purple line and the third-order (qSWIFT-3) with the pink line. The result of the qDRIFTalgorithm (qDRIFT) is plotted with the black line. 
For the deterministic Trotter-Suzuki decomposition, the first-order result (TS-1st) is plotted with the dotted gray line, and the second-order result (TS-2nd) is plotted with the dotted green line. For the randomized Trotter-Suzuki decomposition, the first-order result (RTS-1st) is plotted with the dotted yellow line, and the second-order result (RST-2nd) is plotted with the dotted blue line. }{fig:h}

Fig~\ref{fig:asymptotic} show the asymptotic behaviors of $N$ for each time to achieve the systematic error $\varepsilon = 0.001$, which includes three subfigures corresponding to each molecule: propane with STO-3G basis (left), ethane with 6-31G basis (center), and carbon dioxide with 6-31G basis (right). To generate each molecule Hamiltonian, we use OpenFermion \cite{mcclean2020openfermion}. For each figure, we show the third-order qSWIFT by the pink line (qSWIFT-3), the sixth-order qSWIFT by the pink dotted line (qSWIFT-6), and the qDRIFT by the black line (qDRIFT). 
For the Trotter-Suzuki decomposition, the best of the deterministic method among the first, second, and fourth order is shown by the gray dotted line (TS (Best)), and the best of the randomized method is shown by the green dotted line (RTS (Best)). 

We see that the qSWIFT algorithms outperform the qDRIFT for all $t$ in the sense that the required number of gates $N$ is more than 10 times smaller in the qSWIFT algorithms than in the qDRIFT. While the original qDRIFT reduces the number of gates in the region $t \lesssim 10^8$ from the Trotter-Suzuki decompositions, the qSWIFT algorithms realize a further reduction of the gates. Note that the improvement from the third-order qSWIFT to the sixth-order qSWIFT is not as large as that from the qDRIFT to the third-order qSWIFT. Since there is a trade-off between the order and the number of quantum circuits run as we discuss in Appendix~\ref{section:statistical}, it may be better to utilize the third-order qSWIFT rather than the sixth-order qSWIFT though it depends on the features of quantum devices. 
	
Fig.~\ref{fig:asymptoticPrecise} is the same figure as Fig.~\ref{fig:asymptotic}; other than that, the required systematic error is $\varepsilon = 10^{-6}$. In this case, where more precise time evolution is necessary, the merit of using qSWIFT is much clearer. In qSWIFT-3 (qSWIFT-6), the required number of gates for each time is almost 1,000 (10,000) times smaller than that of qDRIFT. The region where qDRIFT has an advantage over the Trotter-Suzuki is very limited  ($t \lesssim 10^5 \sim 10^6$) due to the bad scaling of qDRIFT in terms of $\varepsilon$. In contrast, the region where qSWIFT has the advantage over Trotter-Suzuki decompositions does not change much from the case of $\varepsilon = 0.001$ ($t \lesssim 10^9 \sim 10^{10}$). 
This result shows the merit of our algorithm; in the case of qDRIFT, we need to increase the number of gates to reduce the systematic error, but in the case of our qSWIFT, it can be reduced just by increasing the order parameter of the algorithm. 

\knadd{
We note that as we write at the end of Section~\ref{section:hs}, we can further improve the bound for the deterministic Trotter-Suzuki decomposition (TS) by using the commutator bounds \cite{childs2021theory} represented as the commutator relation. It should also be noted that the derivation of upper bounds for qDRIFT and qSWIFT involves many triangular inequalities that are loosely bounded. Therefore, the upper bounds for the randomized compiling methods could also be improved by considering relations between operators, presenting a promising direction for future research.}

\subsection{Simulation of the hydrogen molecule}
\label{section:simulation}

We estimate ${\rm Tr}(Q\mathcal{U}(\rho_{\rm init}))$ by using qSWIFT algorithm described in {\bf Algorithm \ref{algorithm:qSWIFT}} with an observable $Q$, time evolution $\mathcal{U}$, and an input state $\rho_{\rm init}$. For $\mathcal{U}$, we implement the time evolution with the hydrogen molecule Hamiltonian with the 6-31G basis and $t=1$. 
Again we use OpenFermion \cite{mcclean2020openfermion} to generate the molecule Hamiltonian. We utilize the Bravyi-Kitaev transformation \cite{bravyi2002fermionic} for transforming the Fermionic operators to Pauli operators. The number of terms in the Hamiltonian $L$ is $184$. The generated Hamiltonian has eight qubits, and therefore, we use a system with nine qubits (including one ancilla qubit). As for the observable $Q$, we choose $Q = ZIIIIIII$, 
and as for the input state, we choose $\rho_{\rm init} = |+\rangle^{\otimes 8}$. 
For comparison, we also estimate  ${\rm Tr}(Q\mathcal{U}(\rho_{\rm init}))$ by using the qDRIFT and the Trotter-Suzuki decomposition. 
As the input parameters of the qSWIFT, we set $N_{\rm shot}^0 = N_{\rm shot}(\vec{n}) = 100$, $N_{\rm sample}^0 = 400,000$, and $N_{\rm sample}(\vec{n}) = C(\vec{n}) \times N_{\rm sample}^0$. For qDRIFT and the randomized Trotter-Suzuki decomposition, we sample $N_{\rm sample}^0$ quantum circuits and perform $N_{\rm shot}^0$ measurements for each circuit. For the deterministic Trotter-Suzuki decomposition, we perform $N_{\rm shot}^0 \times N_{\rm sample}^0$ measurements. For the quantum circuit simulation, we use Qulacs \cite{suzuki2021qulacs}.

Fig.~\ref{fig:h} shows the estimation error of ${\rm Tr}(Q\mathcal{U}(\rho_{\rm init}))$ for each number of gates $N$ for each method. In plotting each point, we perform six trials and show the mean value and the standard deviation of the mean. 
For qSWIFT, we show the result of the second-order (qSWIFT-2) with the purple line and the third-order (qSWIFT-3) with the pink line. The result of the qDRIFT (qDRIFT) is plotted with the black line. 
For the Trotter-Suzuki decomposition, we show the first and second-order results. The minimum $N$ for the first and second Trotter-Suzuki decomposition is $184$ and $368$ ($1840$ for the fourth order). For the deterministic Trotter-Suzuki decomposition, the first-order result (TS-1st) is plotted with the dotted gray line, and the second-order result (TS-2nd) is plotted with the dotted green line. For the randomized Trotter-Suzuki decomposition, the first-order result (RTS-1st) is plotted with the dotted yellow line, and the second-order result (RTS-2nd) is plotted with the dotted blue line.

We see that qSWIFT algorithms outperform the other methods. Particularly, the required $N$ for the third-order qSWIFT for achieving $\varepsilon \sim 0.001$ is almost 10 times smaller than that for qDRIFT, which is consistent with the asymptotic behavior shown in Fig.~\ref{fig:asymptotic}. Consequently, even in the region where Trotter-Suzuki decomposition works better than qDRIFT, qSWIFT algorithms outperform Trotter-Suzuki decompositions.

\section{Conclusion and Discussion}
\label{section:conclusion}
Hamiltonian simulation is a crucial subroutine of various quantum algorithms. Approaches based on the product formulae are practically favored due to their simplicity and ancilla-free nature. There are two representative product formulae-based methods: Trotter-Suzuki decompositions and qDRIFT \cite{campbell2019random}. 
Trotter-Suzuki decompositions have the issue that the number of gates depends on the number of the terms in the Hamiltonian, at least linearly. In contrast, qDRIFT avoids the dependency on the number of terms but has the issue that the number of gates is dependent on the systematic error $\varepsilon$ as $O(1/\varepsilon)$.

In this paper, we propose qSWIFT, a high-order randomized algorithm having both the advantage of the Trotter Suzuki decompositions and qDRIFT, in the sense that its gate count is independent of the number of terms and that the gate count is asymptotically optimal with respect to the precision. We construct the qSWIFT channel and bound the systematic error by the diamond norm. We prove that the qSWIFT channel satisfies the required precision that decreases exponentially with the order parameter in terms of the diamond norm. Then we construct the algorithm by applying the qSWIFT channel to estimate given physical quantities. The algorithm requires a system as simple as qDRIFT; it requires just one ancilla qubit, and only the time evolution operators and the swift operators constructible with the gate in Fig.~\ref{fig:circuitQSWIFT} are necessary to construct the quantum circuits. Our numerical demonstration shows that qSWIFT outperforms qDRIFT with respect to the number of gates for required precision. Particularly when high precision is required, there is a significant advantage of using qSWIFT; the number of gates in the third-order (sixth-order) qSWIFT is 1000 (10000) times smaller than qDRIFT to achieve the systematic error of $\varepsilon =10^{-6}$.

As a future direction, it is beneficial to perform case studies to investigate the performance of qSWIFT in specific problems. Particularly, the literature \cite{lee2021even} points out that the qDRIFT does not perform well in the phase estimation problems, unlike originally expected \cite{campbell2019random} due to the relatively large systematic error for a given number of gates. In contrast, since qSWIFT can successfully reduce systematic error by increasing the order parameter, we expect that the performance in the phase estimation is significantly improved with qSWIFT. 
Also, as we note in Section~\ref{section:introduction}, there are algorithms using many ancilla qubits \cite{BC12,BCC+15,BCK15,LC17,LC19,GSL+19} that achieve a better asymptotic gate scaling with respect to $\lambda$ and $t$ though its constant prefactor is relatively large compared to qDRIFT and qSWIFT. 
Comparing qSWIFT with those algorithms and discussing the advantages of each algorithm in specific problems is a promising direction for future work. 

\knadd{Whether we can improve the scaling with respect to $ \lambda t$ within the framework of qDRIFT and qSWIFT is another important open question. Unlike qDRIFT and qSWIFT, where the number of gates required for a given error scales quadratically with $\lambda t$, the scaling with $t$ can be improved by increasing the order parameter in the Trotter-Suzuki decomposition. Developing methods to enhance the $\lambda t$ scaling would further expedite the realization of practical Hamiltonian simulations.
}

\section*{Acknowledgement}
We thank Nathan Wiebe for helpful discussions. K.N. acknowledges the support of Grant-in-Aid for JSPS Research Fellow 22J01501. A.A.-G.
acknowledges support from the Canada 150 Research Chairs program and CIFAR. A.A.-G. acknowledges the generous support of Anders G. Fr\o seth.

\onecolumngrid
\appendix

\section{Bound for the systematic errors in qSWIFT channels}
\subsection{ Error bound for the second-order qSWIFT channel}
\label{section:secondErrorBound}
In this subsection, we prove the error bound given in Lemma~\ref{lemma:2ndqSWIFTErr} for our second-order qSWIFT channel.
To this end, we utilize the triangle and submultiplicative properties of the diamond norm. Namely, we utilize
\begin{align}	 
\label{eq:triangular}
\diamondnorm{\mathcal{A} + \mathcal{B}}
&\leq 
\diamondnorm{\mathcal{A}} + \diamondnorm{\mathcal{B}},\\
\label{eq:subMulticativity}
\diamondnorm{\mathcal{AB}} &\leq 
\diamondnorm{\mathcal{A}}\diamondnorm{\mathcal{B}}, 
\end{align}
for given channels $\mathcal{A}$ and $\mathcal{B}$.

The diamond distance between the exact time evolution and the qSWIFT channel can be evaluated as
\begin{equation}
\label{eq:errorBound}
\begin{split}	
	\diamondDistance{\mathcal{U}}{\qswift{2}} 
	&= \frac{1}{2}\diamondnorm{
	\mixture_{1,N-1}\left(\Delta_3, \mathcal{E}_N\right)
    +\sum_{k=2}^{N} \mixture_{k,N-k}\left(\Delta_2, \mathcal{E}_N\right)
	} \\
	&\leq \frac{1}{2}\diamondnorm{\mixture_{1,N-1}\left(\Delta_3, \mathcal{E}_N\right)} +
	\frac{1}{2}\sum_{k=2}^N  \diamondnorm{
	\mixture_{k,N-k}\left(\Delta_2, \mathcal{E}_N\right)
	}, \\
	&\leq \frac{1}{2}\sum_{n=3}^{\infty} \frac{\tau^n}{n!}\diamondnorm{M_{1, N-1}(\mathcal{L}^{(n)}, \mathcal{E}_N)} 
	+ \frac{1}{2}\sum_{k=2}^N \sum_{n_1,\cdots n_k=2}^{\infty} \frac{\tau^{\sum_{j}^k n_j}}{n_1! \cdots n_k!}  \diamondnorm{\mixture_{k,N-k}
        \left(
        \left(\mathcal{L}^{(n_1)},\ldots,\mathcal{L}^{(n_k)}\right)
        ,\mathcal{E}_N
        \right)},
\end{split}
\end{equation}
where we use \eqref{eq:triangular} and \eqref{eq:subMulticativity} to show the inequality. 
By the definition of the mixture function $\mixture_{k,N-k}$ in Eq.~\eqref{eq:mixture} and using the triangle and submultiplicative properties of the diamond norm, we obtain
\begin{equation}
\label{eq:boundMixture}
\begin{split}	
 \diamondnorm{\mixture_{k,N-k}
        \left(
        \left(\mathcal{L}^{(n_1)},\ldots,\mathcal{L}^{(n_k)}\right)
        ,\mathcal{E}_N
        \right)} &\leq \sum_{\sigma \in S_{N,k}^{\rm sub}}
 \diamondnorm{\sortingDefault
        \left(
        \left(\mathcal{L}^{(n_1)},\ldots,\mathcal{L}^{(n_k)}\right)
        ,\mathcal{E}_N
        \right)} \\
        & \leq \binom{N}{k} \prod_{j=1}^k \diamondnorm{\mathcal{L}^{(n_j)}} \diamondnorm{\mathcal{E}_N}^{N-k} \\
        & \leq \binom{N}{k} 2^{k+\sum_{j=1}^k n_j}.
\end{split}
\end{equation} 
To show the last inequality we use
\begin{equation}
	\diamondnorm{\mathcal{L}^{(n)}} 
	\leq \diamondnorm{\mathcal{L}}^n + \sum_{\ell=1}^{L}p_\ell 
	\diamondnorm{\mathcal{L}_\ell}^n \leq 2^{n+1},
\end{equation}
which holds since $\diamondnorm{\mathcal{L}} \leq 2$ and $\diamondnorm{\mathcal{L}_\ell} \leq 2$. 

By using \eqref{eq:boundMixture}, 
\begin{equation}
\begin{split}	
	\diamondDistance{\mathcal{U}}{\qswift{2}} 
	&\leq \frac{1}{2}\sum_{n=3}^{\infty} \frac{\tau^n}{n!}\binom{N}{1} 2^{n+1}
	+ 
	\frac{1}{2}\sum_{k=2}^N \sum_{n_1,\cdots n_k=2}^{\infty} \frac{\tau^{\sum_{j}^k n_j}}{n_1! \cdots n_k!}\binom{N}{k} 2^{k+\sum_{j=1}^k n_j} \\
	&=  \frac{1}{2}\sum_{n=3}^{\infty} \frac{N\tau^n}{n!} 2^{n+1} 
	+ \frac{1}{2} \sum_{k=2}^N \sum_{n_1, \cdots n_k=2}^{\xi} \sum_{\xi=4}^{\infty} 
 \updelta\left[\xi, \sum_{j=1}^k n_{j} \right] 
	 \frac{\tau^{\xi}}{n_1! \cdots n_k!}\binom{N}{k} 2^{k+\xi}, 
\end{split}	
\end{equation}
where in the second line, we use 
\begin{equation}
   \sum_{\xi = 4}^{\infty} \updelta\left[\xi, \sum_{j=1}^k n_{j} \right] = 1,
\end{equation}
which holds for a fixed set of integers $\{n_j \}_{j=1}^k$ with $n_{j}\geq 2$ and $k\geq 2$.
By using that $1/n \leq 1/2$ in the summand of the first term and $1/n_j \leq 1/2$ in the summand of the second term, we obtain
\begin{equation}
	\begin{split}
		\diamondDistance{\mathcal{U}}{\qswift{2}} 
	&\leq \frac{1}{2} \sum_{n=3}^{\infty} (2\tau)^n N
		+ \frac{1}{2}  \sum_{\xi=4}^{\infty} 
\sum_{k=2}^{N}  (2\tau)^{\xi}\binom{N}{k}
		 \sum_{n_1, \cdots n_k=2}^{\xi}   \updelta\left[\xi, \sum_{j=1}^k n_{j} \right], \\
	&= 
	 \frac{1}{2} \sum_{n=3}^{\infty} (2\tau)^n N
		+ \frac{1}{2}  \sum_{\xi=4}^{\infty} 
\sum_{k=2}^{\lfloor \xi/2\rfloor}  (2\tau)^{\xi}\binom{N}{k}
		 \sum_{n_1, \cdots n_k=2}^{\xi}   \updelta\left[\xi, \sum_{j=1}^k n_{j} \right], 
	\end{split}
\end{equation}
with $\lfloor x \rfloor$ as the largest integer that does not exceed a real value $x$,
where we use the fact that the summand of the second term vanishes if $k \geq \lfloor \xi/2 \rfloor$. Using
\begin{equation}
\label{eq:deltaBound}
    \sum_{n_1, \cdots, n_k = 2}^{\xi}\updelta\left[\xi, \sum_{j=1}^k n_j\right] 
\leq \xi^k, 
\end{equation}
we obtain
\begin{equation}
	\begin{split}
		\diamondDistance{\mathcal{U}}{\qswift{2}} 
	&\leq \frac{1}{2} \sum_{n=3}^{\infty} (2\tau)^n \binom{N}{1} 
		+ \frac{1}{2}  \sum_{\xi=4}^{\infty} 
\sum_{k=2}^{\lfloor \xi/2\rfloor}  (2\tau)^{\xi}\binom{N}{k}
 \xi^k, \\
	&\leq \frac{1}{2} \sum_{n=3}^{\infty} (2\tau)^n N
	+ \frac{1}{2} \sum_{\xi=4}^{\infty} (2\tau)^{\xi} N^{\lfloor \xi/2 \rfloor}
		\sum_{k=2}^{\lfloor \xi/2 \rfloor} \frac{\xi^k}{k!} \\
	&\leq \frac{1}{2} \sum_{n=3}^{\infty} (2\tau)^n N + \frac{1}{2} \sum_{\xi=4}^{\infty}
	(2e\tau)^{\xi} N^{\lfloor \xi/2 \rfloor} \\
	&\leq  \frac{1}{2} \sum_{\xi=3}^{\infty}
	(2e\tau)^{\xi} N^{\lfloor \xi/2 \rfloor},
	\end{split}
\end{equation}
where to show the third inequality, we use
\begin{equation}
\label{eq:makeExponential}
\sum_{k=1}^{\lfloor \xi/2 \rfloor} \frac{\xi^k}{k !} \leq \sum_{k=0}^{\infty} \frac{\xi^k}{k !} = e^{\xi}.
\end{equation}

Now let us evaluate $\frac{1}{2}\sum_{\xi=2K-1}^{\infty}
	(2e\tau)^{\xi} N^{\lfloor \xi/2 \rfloor}$ (in the current case, $K=2$).   
It can be evaluated depending on if $\xi$ is odd ($\xi = 2m$ with $m$ as an integer) or even ($\xi = 2m-1$) as, 
\begin{equation}
\label{eq:exactSecondBound}
\begin{split}	
\sum_{\xi=2K-1}^{\infty}
	\frac{1}{2}(2e\tau)^{\xi} N^{\lfloor \xi/2 \rfloor} &\leq
\frac{1}{2}\sum_{m=K}^{\infty} (2e \tau )^{2m} N^{m} +  \frac{1}{2}\sum_{m=K}^{\infty} (2e \tau )^{2m-1} N^{m-1}
	\\
	&= \frac{1}{2}\left(1 + \frac{1}{2e N\tau}\right) \sum_{m=K}^{\infty} (2e \tau )^{2m} N^{m} \\
	&= \eta(\lambda t, N) 
	\left( 
		\frac{(2e\lambda t)^2}{N}
	\right)^K, 
\end{split}
\end{equation}
as far as 
\begin{equation}
	\frac{(2e\lambda t)^2}{N} < 1,	
\end{equation}
where 
\begin{equation}
	\eta(x, N) := \frac{1}{2}
\left(1 + \frac{1}{2e x}\right) \frac{1}{1 -  (2e x)^2/N}. 
\end{equation}
By setting $K=2$, we obtain the bound 
\begin{equation}
	\diamondDistance{\mathcal{U}}{\qswift{2}} \leq \eta(\lambda t, N)
	\left( 
		\frac{(2e\lambda t)^2}{N}
	\right)^2.
\end{equation}
In the reasonable parameter range: $1 \leq \lambda t \leq \sqrt{N}/2\sqrt{2}e$ (where the latter inequality holds by choosing suitable $N$), it holds $\eta(\lambda t, N) \leq 3/2$, and therefore
\begin{equation}
	\diamondDistance{\mathcal{U}}{\qswift{2}} \leq \frac{3}{2} \left(
		\frac{(2e\lambda t)^2}{N}
	\right) \in
	\mathcal{O} \left(\left(\frac{(\lambda t)^2}{N}\right)^2\right).
\end{equation}

\subsection{Error bound for the higher-order qSWIFT channels}
\label{section:higherErrorBound}
We now prove the error bound given in Lemma~\ref{lemma:SWIFTErr} for the $K$-th order qSWIFT channel.
By the definition of this channel in Eq.~\eqref{eq:qSWIFTChannel}, we have
\begin{align}
\label{eq:originalHighBound}
\diamondDistance{\mathcal{U}}{\qswift{K}}
&= \frac{1}{2}\diamondnorm{\Phi^{(\infty)} - \qswift{K}}\\
&= \frac{1}{2}
\sum_{\xi=2K-1}^{\infty} \tau^\xi \sum_{k=1}^N \sum_{n_1,\ldots,n_k=2}^\xi 
        \frac{1}{n_1!n_2!\cdots n_k!} 
         \updelta\left[\xi, \sum_{j=1}^k n_j\right]
        \mixture_{k,N-k}
        \left(
        \left(\mathcal{L}^{(n_1)},\ldots,\mathcal{L}^{(n_k)}\right)
        ,\mathcal{E}_N
        \right)\\
& \leq  \frac{1}{2}\sum_{\xi=2K-1}^{\infty} \tau^\xi \sum_{k=1}^N \sum_{n_1,\ldots,n_k=2}^\xi 
        \frac{1}{n_1!n_2!\cdots n_k!} 
         \updelta\left[\xi, \sum_{j=1}^k n_j\right]
        \diamondnorm{\mixture_{k,N-k}
        \left(
        \left(\mathcal{L}^{(n_1)},\ldots,\mathcal{L}^{(n_k)}\right)
        ,\mathcal{E}_N
        \right)},  
\end{align}	
where we used the triangular inequality \eqref{eq:triangular}.

By substituting \eqref{eq:boundMixture} into \eqref{eq:originalHighBound}, we obtain
\begin{equation}
\label{eq:uToPhiK}
\begin{split}	
	\diamondDistance{\mathcal{U}}{\qswift{K}}
	&\leq  \frac{1}{2}\sum_{\xi=2K-1}^{\infty} \tau^\xi \sum_{k=1}^N \sum_{n_1,\ldots,n_k=2}^\xi 
        \frac{1}{n_1!n_2!\cdots n_k!} 
         \updelta\left[\xi, \sum_{j=1}^k n_j\right] 
         \binom{N}{k} 2^{\xi + k} \\
    &\leq \frac{1}{2}\sum_{\xi=2K-1}^{\infty} (2\tau)^\xi \sum_{k=1}^N 
             \binom{N}{k} 
    \sum_{n_1,\ldots,n_k=2}^\xi 
         \updelta\left[\xi, \sum_{j=1}^k n_j\right], \\
    &= \frac{1}{2}\sum_{\xi=2K-1}^{\infty} (2\tau)^{\xi} \sum_{k=1}^{\lfloor \xi/2 \rfloor} \binom{N}{k} \sum_{n_1,\ldots,n_k=2}^\xi 
         \updelta\left[\xi, \sum_{j=1}^k n_j\right]. \\
\end{split}
\end{equation}
We use $1/n_j \leq 1/2$ for ($n_j \geq 2$) in the second inequality and 
to show the third equality, we use the fact that summands with $k > \lfloor \xi/2 \rfloor$ vanishes.
Using \eqref{eq:deltaBound}, we obtain
\begin{equation}
\label{eq:boundLastSum}
\begin{split}	
		\diamondDistance{\mathcal{U}}{\qswift{K}} &\leq\frac{1}{2} \sum_{\xi=2K-1}^{\infty} (2\tau)^{\xi} \sum_{k=1}^{\lfloor \xi/2 \rfloor} \binom{N}{k} \xi^k, \\
		&\leq \frac{1}{2}\sum_{\xi=2K-1}^{\infty} (2\tau)^{\xi} N^{\lfloor \xi/2 \rfloor}\sum_{k=1}^{\lfloor \xi/2 \rfloor} \frac{\xi^k}{k !}, \\
		&\leq \frac{1}{2}\sum_{\xi=2K-1}^{\infty} (2e \tau )^{\xi} N^{\lfloor \xi/2 \rfloor}, 
\end{split}
\end{equation}
where to show the last inequality, we use \eqref{eq:makeExponential}.
By using \eqref{eq:exactSecondBound}, we obtain
\begin{equation}
\label{eq:exactFinalBound}
\begin{split}	
			\diamondDistance{\mathcal{U}}{\qswift{K}} &\leq \eta(\lambda t, N) 
	\left( 
		\frac{(2e\lambda t)^2}{N}
	\right)^K, 
\end{split}
\end{equation}
In the reasonable parameter range: $1 \leq \lambda t \leq \sqrt{N}/2\sqrt{2}e$ again, it holds $\eta(\lambda t, N) \leq 3/2$, and therefore
\begin{equation}
	\diamondDistance{\mathcal{U}}{\qswift{K}} \leq \frac{3}{2} \left(
		\frac{(2e\lambda t)^2}{N}
	\right) \in
	\mathcal{O} \left(\left(\frac{(\lambda t)^2}{N}\right)^K\right).
\end{equation}
Note that when we draw Fig.~\ref{fig:asymptotic} and Fig.~\ref{fig:asymptoticPrecise}, we use the formula \eqref{eq:exactFinalBound}.

\section{Statistical error}	
\label{section:statistical}
Due to the sampling error and the shot noise, there is a statistical error in $\delta \hat{q}^{(k)}(\vec{n})$.
Let us estimate the statistical error in the following. For simplicity, we set $N_{\rm shot}(\vec{n}) = 1$.
Let $\Delta_q\left( (\vec{b_1}, \vec{b_2}, \cdots, \vec{b_k}), \vec{s} \right)$ as the statistical error of $\delta \hat{q}^{(k)}\left(\left(\vec{b_1}, \cdots \vec{b_k}\right), \vec{s} \right)$. Then, from the central limit theorem, it behaves as
\begin{equation}
	 \Delta_q\left( (\vec{b_1}, \vec{b_2}, \cdots, \vec{b_k}), \vec{s} \right)
	 \sim
	 \frac{{\rm Var}\left(\delta \hat{q}^{(k)}\left(\left(\vec{b_1}, \cdots \vec{b_k}\right), \vec{s} \right)\right)}{\sqrt{N_{\rm sample}(\vec{n})}}, 
\end{equation}
where ${\rm Var}(A)$ denotes the variance of the variable $A$. In the rest of the section, we use `$\sim$' with the same meaning. 
Since 
\begin{equation}
	-1 \leq \delta \hat{q}^{(k)}\left(\left(\vec{b_1}, \cdots \vec{b_k}\right), \vec{s} \right) \leq 1, 
\end{equation}
from Popoviciu's inequality on variances, it holds ${\rm Var}\left(\delta \hat{q}^{(k)}\left(\left(\vec{b_1}, \cdots \vec{b_k}\right), \vec{s} \right)\right) \leq 1$, and 
\begin{equation}
		 \Delta_q \left( (\vec{b_1}, \vec{b_2}, \cdots, \vec{b_k}), \vec{s} \right)
	 \lesssim \frac{1}{\sqrt{N_{\rm sample}(\vec{n})}}.
\end{equation}
Using the fact that each $\delta \hat{q}^{(k)}\left(\left(\vec{b_1}, \cdots \vec{b_k}\right), \vec{s} \right)$ is independent, we can estimate the statistical error of $\delta \hat{q}^{(k)}(\vec{n})$ as 
\begin{equation}
\begin{split}	
		|\delta q^{(k)}(\vec{n}) - \delta \hat{q}^{(k)}(\vec{n})| &\leq c^{(k)}(\vec{n}) 
		\sqrt{
					 \sum_{\vec{b_1}\cdots \vec{b_k}} \sum_{\vec{s}}\left[ \Delta_q\left( (\vec{b_1}, \vec{b_2}, \cdots, \vec{b_k}), \vec{s} \right)\right]^2
		} \\
		&\lesssim c^{(k)}(\vec{n}) \sqrt{\frac{2^{\sum_j n_j + k}}{N_{\rm sample}(\vec{n})}}, 
\end{split}
\end{equation}
where to show the second inequality, we use $ \sum_{\vec{b_1}\cdots \vec{b_k}} \sum_{\vec{s}} 1 = 2^{\sum_j n_j + k}$. In other words, the statistical error is bounded as 
\begin{equation}
	|\delta q^{(k)}(\vec{n}) - \delta \hat{q}^{(k)}(\vec{n})| \leq \varepsilon, 
\end{equation}
if we set 
\begin{equation}
	N_{\rm sample}(\vec{n}) \sim \frac{[c^{(k)}(\vec{n})]^2 \times 2^{\sum_j n_j + k}}{\varepsilon^2}.
\end{equation}
Let $N_{\rm total}^{(k)}(\vec{n})$ be the total circuit run for calculating $\delta \hat{q}^{(k)}(\vec{n})$. Then it holds 
\begin{equation}
\label{eq:sample-bound}
	N_{\rm total}(\vec{n}) = 2^{\sum_j n_j + k} \times N_{\rm sample}(\vec{n}) \sim \frac{[c^{(k)}(\vec{n})]^2 \times 2^{2\sum_j n_j + 2k}}{\varepsilon^2}.
\end{equation}

Let us further clarify the implication of \eqref{eq:sample-bound}.
Recall that in the qSWIFT algorithm, we only compute $\delta \hat{q}^{(k)}(\vec{n})$ only if 
\begin{equation}
\label{eq:sumConstraint}
\sum_{j=1}^k n_j = \xi,~~n_j \geq 2~(\forall j).
\end{equation}
When the condition \eqref{eq:sumConstraint} is satisfied, it holds
\begin{equation}
	c^{(k)}(\vec{n}) \leq N^k \left(\frac{\tau^\xi}{2^k} \right), 
\end{equation}
where $\xi \geq 2k$ (the equality holds when $n_j = 2$ for all $j$). Conversely, 
\begin{equation}
	c^{(k)}(\vec{n}) \leq \left\{
	\begin{array}{cc}
	\left(\frac{(\lambda t)^2}{2N}\right)^{\frac{\xi}{2}} & \text{if }n_j = 2 \text{ for all } j\\
	\sqrt{\frac{2}{N}}\left(\frac{(\lambda t)^2}{2N}\right)^{\frac{\xi}{2}} & \text{otherwise}
	\end{array}
	\right. ,
\end{equation}
meaning that $c^{(k)}(\vec{n})$ is suppressed by the factor $\sqrt{2/N}$ unless $n_j=2$ is satisfied for all $j$.
Therefore, the dominant source of the quantum circuits run comes from the calculation of $\delta \hat{q}^{(k)}(\vec{n})$ with $n_j = 2$ for all $j$; and the number of measurements for calculating $\delta \hat{q}^{(k)}(\vec{n})$ with other $\vec{n}$ asymptotically becomes negligible as $N$ becomes large. 
In the asymptotic limit, the total number of quantum circuits run $N_{\rm total}$ for computing all terms in $q^{(K)}$ within the error $\varepsilon$ in $2K$-th order qSWIFT can be estimated as
\begin{align}
	N_{\rm total} &\approx N_{\rm sample}^0 + N_{\rm total}^{(1)}\left(\{2\}\right) + N_{\rm total}^{(2)}\left(\{2, 2\}\right) + \cdots N_{\rm total}^{(K)}\left(\{2, \cdots 2\}\right) \\
	&\sim  \frac{1}{\varepsilon^2}\sum_{k=0}^K  \left(\frac{(2\lambda t)^2}{N}\right)^{\xi}, 
\end{align}
which includes the number of samples $N_{\rm sample}^0$ to compute $q^{(1)}$.
The approximation in the first line denotes the asymptotic limit. 
To obtain the last expression, we use \eqref{eq:sample-bound} and also use that if $N_{\rm shot}^0 = 1$, required number of samples to reduce the statistical error of $q^{(1)}$ within $\varepsilon$ is
\begin{equation}
	N_{\rm sample}^0 \sim \frac{1}{\varepsilon^2}
\end{equation}
from the central limit theorem. To reduce the statistical error of $q^{(K)}$ less than $\varepsilon_{\rm total}$, we should set $\varepsilon = \varepsilon_{\rm total}/\sqrt{K+1}$, and therefore, 
\begin{equation}
	N_{\rm total} \sim \frac{K+1}{\varepsilon_{\rm total}^2} \sum_{k=0}^K  \left(\frac{(2\lambda t)^2}{N}\right)^{\xi}.
\end{equation}
Thus, if we fix $\varepsilon_{\rm total}$, $N_{\rm total}$ scales less than quadratically with $K$ as far as $(2\lambda t)^2/N < 1$.

\section{All-order qSWIFT}
\label{section:allOrderqSWIFT}
In the main text, we constructed the second-order and higher-order versions of qSWIFT, which include systematic errors dependent on the order parameter. In this section, we demonstrate that we can create an all-order version of qSWIFT that has no systematic error.

The expectation value of an observable can be expanded as 
\begin{equation}
\begin{split}	
	{\rm Tr}\left(Q\mathcal{U} (\rho_{\rm init})\right) 
	&= {\rm Tr}\left(Q(\mathcal{E}_N + \Delta_2)^N(\rho_{\rm init})\right) \\
	&= {\rm Tr}\left(Q\left(\mathcal{E}_N + \sum_{n=2}^{\infty} \frac{\tau^n}{n!}\mathcal{L}^{(n)}\right)^N(\rho_{\rm init})\right) \\
	&= {\rm Tr}\left(Q\left(\mathcal{E}_N + \sum_{n=2}^{\infty} \frac{\tau^n}{n!}
	\sum_{\vec{\ell}} \sum_{s=0}^{1} (-1)^s P_{s}^{(n)}(\vec{\ell}) \mathcal{L}_n(\vec{\ell})\right)^N(\rho_{\rm init})\right) \\
	&= {\rm Tr}\left(\tilde{Q}\left(\tilde{\mathcal{E}}_N + \sum_{n=2}^{\infty} \frac{\tau^n}{n!}
	\sum_{\vec{\ell}} \sum_{s=0}^{1} (-1)^s P_{s}^{(n)}(\vec{\ell}) \tilde{\mathcal{L}}_n(\vec{\ell})\right)^N(\tilde{\rho}_{\rm init})\right) \\
	&= {\rm Tr}\left(\tilde{Q}\tilde{\mathcal{U}}^N_N (\tilde{\rho}_{\rm init})\right), 
\end{split}
\end{equation}
where we use \eqref{eq:Deltak} in the second equality, and we use \eqref{eq:d-expand} and \eqref{eq:lNExpand} in the third equality. We can readily show the fourth equality by using \eqref{eq:higherOrderTransform}.
In the last equality, we define
\begin{equation}
		\tilde{\mathcal{U}}_N := \tilde{\mathcal{E}}_N +
	 \sum_{n=2}^{\infty} \frac{\tau^n}{n!} \sum_{s=0}^1 \sum_{\vec{b} \in \{0, 1\}^{\otimes n} } \sum_{\vec{\ell}}
	 (-1)^s P_s^{(n)}(\vec{\ell}) \tilde{S}_n^{(b)}(\vec{\ell}).
\end{equation}
We can rewrite $\tilde{\mathcal{U}}_N$ by using the physical channel as follows:
\begin{equation}
 \label{eq:allOrderExpansion}
\begin{split}	
	\tilde{\mathcal{U}}_N
	 &=  \tilde{\mathcal{E}}_N +
	 \sum_{n=2}^{\infty} \frac{2^{n+1}\tau^n}{n!} \frac{1}{2}\sum_{s=0}^1 \frac{1}{2^n}\sum_{\vec{b} \in \{0, 1\}^{\otimes n} } \sum_{\vec{\ell}}
	 (-1)^s P_s^{(n)}(\vec{\ell}) \tilde{S}_n^{(b)}(\vec{\ell}) \\
	 &= \tilde{\mathcal{E}}_N + \sum_{n=2}^{\infty} \beta(n) \tilde{\mathcal{W}}_n \\
	 &= B \mathcal{E}^{\rm (all)}_N.
\end{split}
\end{equation}
In the second equality of \eqref{eq:allOrderExpansion}, we define
\begin{equation}
\beta(n) := \frac{2^{n+1}\tau^n}{n!},~\tilde{\mathcal{W}}_n :=  \frac{1}{2}\sum_{s=0}^1 \frac{1}{2^n}\sum_{\vec{b} \in \{0, 1\}^{\otimes n} } \sum_{\vec{\ell}}
	 (-1)^s P_s^{(n)}(\vec{\ell}) \tilde{S}_n^{(b)}(\vec{\ell}), 
\end{equation}
where $\tilde{\mathcal{W}}_n (n\geq 2)$ is the physical channel; we can implement the process by sampling $s$ and $\vec{b}$ uniformly, sampling $\vec{\ell}$ according to $P_s^{(n)}(\vec{\ell})$, and applying $(-1)^s \tilde{S}_n^{(b)}(\vec{\ell})$.
In the last equality, we define 
\begin{equation}
	B := 1 + \sum_{n=2}^{\infty} \beta(n) = e^{(2\ln 2) \tau} - 4\tau  - 1,
\end{equation}
and the channel $\mathcal{E}^{(\rm all)}_N$, the physical channel implemented by applying $\mathcal{E_N}$ with the probability $1/B$ and $\mathcal{W}_n (n\geq 2)$ with the probability $\beta(n)/B$. 
Consequently, we obtain
\begin{equation}
		{\rm Tr}\left(Q\mathcal{U} (\rho_{\rm init})\right) 	= B^N {\rm Tr}\left(\tilde{Q}\left(\mathcal{E}^{(\rm all)}_N\right)^N (\tilde{\rho}_{\rm init})\right). 
\end{equation}
We see that there is no systematic error on the right-hand side. For a given number of samples, denoted as $N_{\rm sample}$, the statistical error $\epsilon_{\rm st}$ scales as 
\begin{equation}	
\epsilon_{\rm st} \in \mathcal{O}\left(\frac{B^{N}}{\sqrt{N_{\rm sample}}}\right).
\end{equation}
Since $B^N < e^{(2\ln 2) (\lambda t)^2/N}$, we get $N_{\rm sample} \in \mathcal{O}(1/\epsilon^2_{\rm st})$ by setting $N \in \mathcal{O}\left( (\lambda t)^2\right)$.

We note that even though \( N \) is upper-bounded, there is no theoretical upper bound on the number of gates, as the count of swift operators in \( \tilde{\mathcal{W}}_n \) is determined by \( n \). Here, \( n \) is not bounded and can take on a large value with probability $\beta(n)/B$.  Conversely, the count of swift operators in the second- or higher-order qSWIFT is upper-bounded by the order parameter, and therefore, the number of gates is also upper-bounded. Thus, in practical situations where the number of operational gates is limited, the second- or higher-order qSWIFT may be more advantageous than the previously introduced all-order qSWIFT.


\section{Note on the LCU-based randomized approach}	
\label{section:lcuBased}
The literature \cite{wan2022randomized} provides a higher-order randomized method for phase estimation. The calculation includes the evaluation of 
$
{\rm Tr}\left[\rho e^{iHt} \right]
$
with $H = \sum_{\ell=1}^L h_{\ell} H_{\ell}$, where again we define $\{H_{\ell}\}_{\ell=1}^L$ so that $h_{\ell} > 0$ and define $\lambda := \sum_{\ell} h_{\ell}$.  For that, they propose an all-order randomized method for estimating ${\rm Tr}\left[\rho e^{iHt} \right]$ based on the linear combination of the unitary (LCU) approach. In this section, we show that by extending the randomized method, we can construct another all-order randomized method for estimating ${\rm Tr}(Q e^{\upi Ht} \rho e^{-\upi Ht})$, which is the target of our qSWIFT algorithm.

We begin by reviewing the method given in \cite{wan2022randomized} and discuss how to extend this method to estimate the expectation value of an observable. Then, we discuss the difference between the LCU-based approach and the all-order qSWIFT introduced in Appendix~\ref{section:allOrderqSWIFT}. 

\newcommand{\qmlcu}[0]{q_m}
\newcommand{\cmlcu}[0]{c_m}
\newcommand{\constlcu}[0]{C}

\subsubsection{The LCU-based
randomized approach}
To estimate ${\rm Tr}[\rho e^{\upi Ht}]$, the authors of \cite{wan2022randomized} utilize the expansion
\begin{equation} \label{cmVm}
e^{i H t/N} = \sum_m \cmlcu W_m\,,
\end{equation}
where 
\begin{equation}
	W_m \rightarrow (\upi~{\rm sign}(t))^n H_{\ell_1}\dots H_{\ell_n} V_{\ell'}^{(n)},~~c_m \rightarrow \frac{1}{n!} \left(\frac{\lambda |t|}{N}\right)^{n}\sqrt{1 + \left(\frac{\lambda t}{N(n+1)}\right)^2}p_{\ell_1} \dots p_{\ell_n} p_{\ell'} > 0, 
\end{equation}
with 
\begin{align} \label{eq:xxx}
V_{\ell'}^{(n)} = \exp(\upi\theta_n H_{\ell'}), \quad\text{with }\theta_n \coloneqq \arccos\left(\left[1 + \left(\frac{\lambda t}{N(n+1)}\right)^2 \right]^{-1/2}\right).
\end{align}
The multi-index $m$ denotes the indices $(n, \vec{\ell}, \ell^{\prime})$.
Then 
\begin{align} 
\label{eq:lcuExpansion}
e^{iHt} &= \sum_{m_1,\dots, m_N} c_{m_1}\dots c_{m_N} W_{m_1}\dots W_{m_N} = \constlcu^{N} \sum_{m_1,\dots, m_N} q_{m_1}\dots q_{m_N} W_{m_1}\dots W_{m_N}, 
\end{align}
where $\qmlcu = \cmlcu / \constlcu$ with $C= \sum_{m} c_m$. Since $\qmlcu > 0$ and $\sum_m \qmlcu = 1$, $\{\qmlcu\}$ can be interpreted as a probability distribution. 

By using the expansion, we obtain 
\begin{equation}
\label{eq:lcuEstimation}
	{\rm Tr}\left[ \rho e^{iHt} \right] = C^N  \sum_{m_1,\dots, m_N}
	q_{m_1}\dots q_{m_N} \left(
	\real\left[{\rm Tr}\left(\rho W_{m_1}\dots W_{m_N} \right)\right] + 
	\upi \imaginary\left[{\rm Tr}\left(\rho W_{m_1}\dots W_{m_N} \right)\right] 
	\right).
\end{equation}
We can estimate the right-hand side, by sampling $(m_1 \dots m_N)$ according to $q_{m_1} \dots q_{m_N}$ and evaluate the real part and the imaginary part of ${\rm Tr}\left(\rho W_{m_1}\dots W_{m_N} \right)$ by the Hadamard test repeatedly. There is no systematic error. The statistical error $\mathcal{\epsilon_{\rm st}}$ scales as 
\begin{equation}
	\mathcal{\epsilon_{\rm st}} \in \mathcal{O}\left(\frac{C^N}{\sqrt{N_{\rm sample}}}\right)
\end{equation}
with $N_{\rm sample}$ as the number of sampling the set $(m_1, \dots m_N)$. In \cite{wan2022randomized}, $C \in \mathcal{O}\left(\exp\left( (\lambda t)^2/N^2\right)\right)$, i.e., $C^N = \mathcal{O}\left(\exp\left((\lambda t)^2/N \right)\right)$ in our notation. Thus, we need $N = O\left((\lambda t)^2\right)$ so that
$N_{\rm sample} \in \mathcal{O}\left(1/\epsilon_{\rm st}^2 \right)$. 

\subsubsection{Application to the Hamiltonian simulation problem}
We can utilize the expansion \eqref{eq:lcuExpansion} to calculate the expectation value after applying the time evolution:
\begin{align}
	{\rm Tr}\left(Q e^{iHt} \rho e^{-iHt}\right) 
	&= C^{2N} \sum_{m_1^{\prime}\dots m_N^{\prime}} \sum_{m_1\dots m_N}
		q_{m_1^{\prime}}\dots q_{m_N^{\prime}} 
		q_{m_1}\dots q_{m_N} 
		{\rm Tr} \left( Q
		W_{m_1^{\prime}}\dots W_{m_N^{\prime}}
		\rho 
		W_{m_1}^{\dagger} \dots W_{m_N}^{\dagger}
		\right) \\
	&= C^{2N} \sum_{m_1^{\prime}\dots m_N^{\prime}} \sum_{m_1\dots m_N}
		q_{m_1^{\prime}}\dots q_{m_N^{\prime}} 
		q_{m_1}\dots q_{m_N} 
		\real{\left[
		{\rm Tr} \left( Q
		W_{m_1^{\prime}}\dots W_{m_N^{\prime}}
		\rho 
		W_{m_1}^{\dagger} \dots W_{m_N}^{\dagger}
		\right)\right]},  
		\label{eq:lcuObservable}
\end{align}
where we use that the left-hand side is the real value in the second equality. 
As in the case of \eqref{eq:lcuEstimation}, we can estimate the value of \eqref{eq:lcuObservable}, by sampling $(m_1, \dots m_N)$ and $(m_1^{\prime}, \dots m_N^{\prime})$ and estimating $\real{\left[
		{\rm Tr} \left( Q
		W_{m_1^{\prime}}\dots W_{m_N^{\prime}}
		\rho 
		W_{m_1}^{\dagger} \dots W_{m_N}^{\dagger}
		\right)\right]}$ by the Hadamard test. Again, there is no systematic error, and by setting $N=O((\lambda t)^2)$, the number of samples $N_{\rm sample}$ scales as $N_{\rm sample} \in \mathcal{O}(1/\epsilon_{\rm st}^2)$. Therefore, the behavior of both the systematic error and the statistical error with respect to \( N \) is the same as that of the all-order qSWIFT introduced in Appendix~\ref{section:allOrderqSWIFT}.

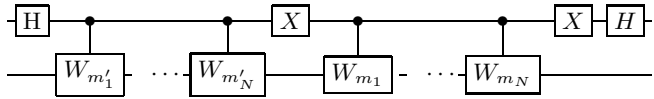
\begin{figure}[ht]
 ~\Qcircuit @C=.3em @R=.7em { & \gate{{\rm H}} & \ctrl{1} & \qw & \qw & \qw & \qw & \qw & \qw & \qw & 
\qw & \ctrl{1} & \gate{X} & \qw & \ctrl{1} & \qw 
&\qw & \qw & \qw  & \qw & \qw & \qw & \qw 
& \qw &  \ctrl{1} & \qw & \gate{X} & \qw & \gate{H} & \qw &
 \\
&\qw & \gate{W_{m_1^{\prime}} } & \qw &
\rstick{\cdots} & \lstick{} & \lstick{} & \lstick{} & \lstick{} & \lstick{} & \lstick{} & \gate{W_{m_N^{\prime}} } & \qw 
&
\qw
&
\gate{W_{m_1}}
&
 & \qw &\rstick{\cdots} & \lstick{} & \lstick{} & \lstick{} &  
\lstick{} & \lstick{} & \lstick{} &\gate{W_{m_N}} & \qw & \qw & \qw & \qw & \qw 
}
\caption{Quantum circuits for evaluating \eqref{eq:lcuObservable} by the LCU-based approach.}
\label{fig:circuitLCUExpansion}
\end{figure}

However, the LCU-based method and our qSWIFT approach have a key distinction: the former expands the unitary operation, while the latter expands the unitary channel. This difference in expansion methodology results in distinct quantum circuits. In the LCU-based approach, the Hadamard test is necessary for calculating the real part of the trace function, requiring the implementation of controlled-$W_m$ operations. For instance, the quantum circuit for evaluating \eqref{eq:lcuObservable} is depicted in Fig.~\ref{fig:circuitLCUExpansion} (we note that the circuit is also used in another higher-order randomization protocol \cite{faehrmann2022randomizing}). Given that each $W_m$ operation involves a rotational operator of the form $e^{\upi H_{\ell} t^{\prime}}$ (with $t^{\prime}$ as a real value), a total of $2N$ controlled-$e^{\upi H_{\ell} t^{\prime}}$ operations are needed for the circuit. Conversely, in the qSWIFT approach, as illustrated in Fig.~\ref{fig:circuitQSWIFTALL}, no controlled-$e^{\upi H_{\ell} t^{\prime}}$ operations are necessary. The only interactions with the ancilla qubit involve swift operators, as all exponential time evolution operators are encapsulated within the qDRIFT channel, which can be implemented without the need for an ancilla qubit. 

The implementation of the controlled-$e^{\upi H_{\ell} t^{\prime}}$ operation requires additional controlled gates in the LCU-based methods compared to the implementation of $e^{\upi H_{\ell} t^{\prime}}$ operation. In particular, in the simulation where one body term is dominant, the LCU-based method needs many more controlled gates than in qSWIFT. To demonstrate this, let us consider the following Hamiltonian, written as the summation of the multi-body terms (the first summation) and the one-body terms (the second summation):
\begin{equation}
	H = \frac{J}{h^{\prime}_{\rm sum} n} \sum_{\ell=1}^{L^{\prime}} h_{\ell}^{\prime} P_{\ell} + \frac{h}{c_{\rm sum}} \sum_{j=1}^n \sum_{k=x,y,z} c_j^k \sigma_j^k.
\label{eq:oneBodyDominant}
\end{equation}
The first summation corresponds to $L^{\prime}$ multi-body terms, where $P_{\ell}$ is the tensor product of the Pauli operators that act non-trivially on multiple qubits, 
with $h_{\ell}^{\prime}$ as a real coefficient, $h_{\rm sum}^{\prime} := \sum_{\ell} |h_{\ell}^{\prime}|$, and $J$ is a positive value. 
The second summation corresponds to $3n$ one-body terms, with
$n$ as the number of qubits, $\sigma_j^k$ as one of the Pauli operators acting on the $j$-th qubit, $c_j^k$ and $h_{\ell}^{\prime}$ as real coefficients, $c_{\rm sum} = \sum_{j=1}^b \sum_{k=x,y,z} |c_j^k|$ and $h$ as a positive value. We assume the case where one-body term is dominant in the sense that $J \ll h$. The sum of all absolute values of the coefficients is given as $\lambda = J/n + h$.

Our objective of the simulation is approximately computing ${\rm Tr}\left(Q e^{\upi  Ht} \rho e^{-\upi Ht}\right)$ using directly implementable time evolutions: $e^{iP_{\ell} t^{\prime}}$ and $e^{i\sigma_j^k t^{\prime}}$. 
Note that the collective neutrino oscillation problem is an example of the Hamiltonian simulation where the one body term is dominant and has been already discussed in \cite{rajput2022hybridized}.
On the one hand, in the LCU-based approach, the required number of controlled time evolution gates is $O((\lambda t)^2)$, and since each controlled time evolution requires at least one controlled gate, the total number of controlled gates is also $O((\lambda t)^2)$. On the other hand, the number of time evolution gates is also $O((\lambda t)^2)$ in the all-order qSWIFT. However, most of the time evolution gates sampled are $e^{i\sigma_j^k t^{\prime}}$, which do not need the controlled gate; the number of $e^{iP_{\ell} t^{\prime}}$ sampled is $O(J/(\lambda n))$ on average. Since each $e^{iP_{\ell} t^{\prime}}$ requires at most $O(n)$ controlled gates, the total number of controlled gates in qSWIFT is at most
\begin{equation}
	\mathcal{O} \left((\lambda t)^2 \times \frac{J}{\lambda n} \times n\right)
	\sim \mathcal{O}\left((\lambda t)^2 \times \frac{J}{h}\right), 
\end{equation}
where to show the expression on the right-hand side, we use $\lambda \sim h$ since $J \ll h$. Therefore, for the simulation with the Hamiltonian \eqref{eq:oneBodyDominant}, qSWIFT achieves a significant reduction of the CNOT gates by the factor $J/h$.

\bibliography{main}

\end{document}